\documentclass[lettersize,journal]{IEEEtran}
\usepackage{amsmath,amsfonts}
\usepackage{algorithmic}
\usepackage{algorithm}
\usepackage{array}
\usepackage[caption=false,font=normalsize,labelfont=sf,textfont=sf]{subfig}
\usepackage{textcomp}
\usepackage{xcolor}
\usepackage{verbatim}
\usepackage{graphicx}
\usepackage{cite}
\graphicspath{{TPEimages/}{TPEbio/}}
\usepackage{geometry}
\usepackage{cite}
\usepackage{picinpar}
\usepackage{amsmath}
\usepackage{lipsum}  
\usepackage{booktabs}
\usepackage{amssymb}
\usepackage{mathcomp}
\usepackage{empheq}
\usepackage{enumitem} 
\usepackage{amsmath,amssymb,mathtools}
\usepackage{multirow}

\usepackage{makecell}
\usepackage[table]{xcolor}
\renewcommand{\arraystretch}{1} % 矩阵之间的行距

\makeatletter
\def\@listi{\leftmargin\leftmargini
	\parsep 0pt
	\topsep 0pt
	\itemsep 0pt}
\let\@listI\@listi
\makeatother
\usepackage{enumitem}
\setlist[itemize]{noitemsep, topsep=0pt, leftmargin=*} % leftmargin=* 去掉缩进

\begin{document}
	
	\title{Full-Time-Scale Power Management for Hybrid AC/DC/DS Microgrid with Dynamic Concatenation and Global Equivalent Circuit Modeling}
	
	\author{
		Pengfeng Lin,~\IEEEmembership{Senior Member,~IEEE,}
		Qingzuo Meng,
		Yujie Wang,~\IEEEmembership{Student Member,~IEEE,}
		Miao Zhu, ~\IEEEmembership{Senior Member,~IEEE,}
		Amer M. Ghias, ~\IEEEmembership{Senior Member,~IEEE,}
		Syed M. Islam, ~\IEEEmembership{Fellow,~IEEE,}
		%Changyun Wen, ~\IEEEmembership{Fellow,~IEEE,}
		Frede Blaabjerg, ~\IEEEmembership{Fellow,~IEEE}

		% <-this % stops a space
		%\thanks{Pengfeng Lin, Qingzuo Meng, Yujie Wang, and Miao Zhu are with Shanghai Jiao Tong University, China. (Email: linpengfeng@ieee.org, \{rexxxx, xiaowangtongxue, miaozhu\} @sjtu.edu.cn)
			
		%Amer M. Ghias is with Nanyang Technological University, Singapore. (Email: amer.ghias@ntu.edu.sg)
		
		%Syed Islam is with Federation University, Australia. (Email: s.islam@federation.edu.au)
			
		% Frede Blaabjerg is with Aalborg University, Denmark. (Email: fbl@energy.aau.dk)}% <-this % stops a space
		%\thanks{Manuscript received April 19, 2021; revised August 16, 2021.}
		\thanks{Manuscript received on 23 May, 2026; revised on 24 Aug, 2026 and 6 Nov, 2026, respectively; accepted on 14 Nov, 2026. This work is supported by Marie Skłodowska-Curie Actions Fellowship with Project ID 101268616-MRSSG and the National Natural Science Foundation of China under Grant No.52507124. (Corresponding author: Qingzuo Meng).
			
			Pengfeng Lin is with University of Cambridge, UK, and Shanghai Jiao Tong University, China. (Email: linpengfeng@ieee.org)
			
			Qingzuo Meng is with Electric Power Dispatching and Control Center of Guizhou Power Grid Co., Ltd. (Email: rexxxx@sjtu.edu.cn)
			
			Yujie Wang and Miao Zhu are with Shanghai Jiao Tong University, China. (Email: \{xiaowangtongxue, miaozhu\} @sjtu.edu.cn).
			
			Amer M. Ghias is with The University of Glasgow, UK. (Email: Amer.Ghias@glasgow.ac.uk).
			
			Syed M. Islam is with Federation University, Australia. (Email: s.islam@federation.edu.au)
			
			Frede Blaabjerg is with Aalborg University, Denmark. (Email: fbl@energy.aau.dk).
			
		}
		
	}

%	\markboth{Journal of \LaTeX\ Class Files,~Vol.~14, No.~8, August~2021}%
%	{Shell \MakeLowercase{\textit{et al.}}: A Sample Article Using IEEEtran.cls for IEEE Journals}

	\maketitle

	\IEEEpubidadjcol
	\vspace{1.2\baselineskip}  % 如果发现仍重叠可以调节为 1.5 或 2
	
	\begin{abstract}
		Hybrid AC/DC/distributed energy storage (DS) microgrids (MGs) provide an effective solution for improving power accessibility and supply reliability in remote and isolated areas. 
		Conventional static power management aims to proportionally allocate power based on unit ratings by scheduling the operation of interlinking converters (ILCs). Recent studies on transient power management emphasize inertia coupling among subgrids by dynamically regulating ILC power flows, thereby enhancing the system resilience against disturbances. In existing literature, these two objectives are normally addressed in a separate fashion and a unified scheme to simultaneously achieve them is rarely found. Furthermore, conventional static management neglects droop-induced frequency and voltage deviations which may seriously compromise the safe operation of sensitive loads. To address these limitations, in this paper, we proposes a full-timescale (FTS) power management for the hybrid MG is proposed. 
		Autonomous frequency/voltage restoration controls are deployed in each subgrid to eliminate droop-induced frequency and voltage deviations, based on which a novel FTS dynamic concatenator is proposed to chain up the static power sharing and transient inertia sharing across different timescales. Moreover, a novel global equivalent circuit model (GECM) is presented to interpret the holistic hybrid system from electrical circuit perspective. Then the static and transient performance of the complex hybrid AC/DC/DS MG can be quantitatively characterized to facilitate easier engineering practices. In-house experiments validate that the proposed FTS dynamic concatenator ensures static power sharing and transient inertia sharing through the ILC in full timescales. 
		\end{abstract}
	
	
	\section{Introduction }
	\subsection{Research Background}
	Conventional power grids endure comparatively higher upfront costs and operating costs when they have to be extended and constructed in remote regions \cite{YangFang, Zia}. According to reports released by the International Energy Agency (IEA), due to limited access to main grids, there are still around 65 millions of people living on isolated islands in countries of Southeast Asia who cannot use stable electricity. To meet basic needs of using electricity in these people, microgrids (MGs) can be built on islands spread widely in Southeast Asia regions and operated in island modes. In recent years, hybrid AC / DC / distributed energy storage (DS) MGs have gained increased attention as an effective paradigm to uniformly integrate energy storage systems into AC and DC subgrids \cite{Blaabjerg}. 
	
	A typical hybrid AC/DC/DS MG is illustrated in Fig. 1 and comprises four principal components: an AC subgrid, a DC subgrid, a DS subgrid, and an interlinking converter (ILC) that coordinates the power flows across subgrids. By integrating renewable energy sources (RES) such as photovoltaic (PV), wind, and small-scale hydro units into appropriate AC or DC subgrids, the hybrid MG can harness locally available resources with lower operational costs compared to pure AC or DC MGs. The DS subgrid further provides buffering and load balancing capabilities to mitigate the intermittency of RESs. In addition, ILC enables flexible power exchanges and coordinated dispatch for the entire hybrid MG, which is crucial to maintain the stability and reliability of the system \cite{JinChi2014,ZhangZhe2021}.
	
	\begin{figure*}[t]
		\setlength{\abovecaptionskip}{-3pt}
		\setlength{\belowcaptionskip}{-5pt}
		\centering
		\includegraphics[scale=1]{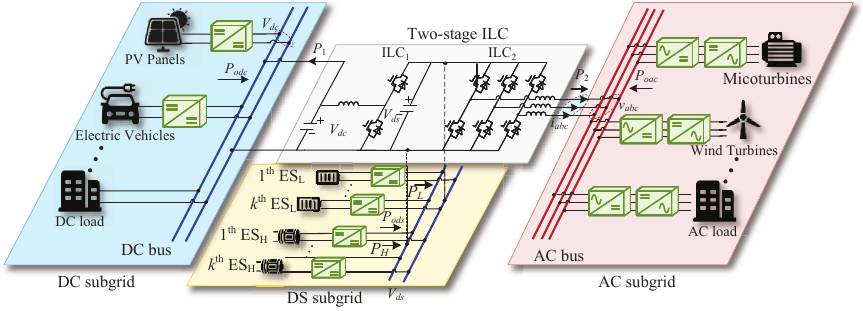}
		\caption{ System configuration of a typical hybrid AC/DC/DS MG.}
		\label{Fig. AC/DC/DS MG}
		\vspace{-1.5em}
	\end{figure*}

		\begin{figure}[t]
		\setlength{\abovecaptionskip}{-3pt}
		\setlength{\belowcaptionskip}{-5pt}
		\centering
		\includegraphics[scale=1]{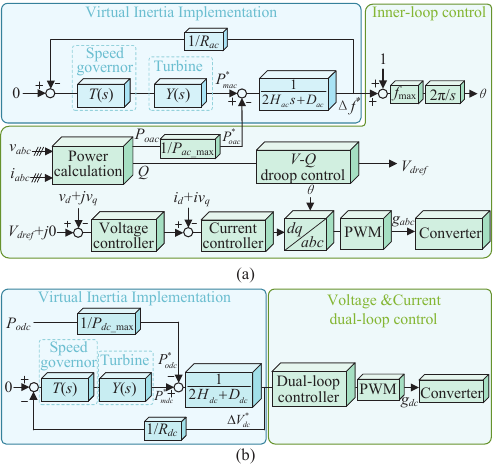}
		\caption{Control diagrams of AC and DC subgrids. (a) DC subgrid. (b) AC subgrid.}
		\label{Fig. ac and dc control}
		\vspace{-0.5em}
	\end{figure}

	\begin{figure}[t]
		\setlength{\abovecaptionskip}{-3pt}
		\setlength{\belowcaptionskip}{-5pt}
		\centering
		\includegraphics[scale=1]{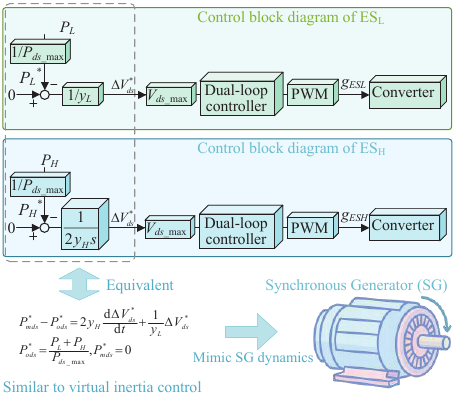}
		\caption{Control block diagrams of $\rm{E{S_L}}$ and $\rm{E{S_H}}$ and their analogy with the swing equation of a synchronous generator.}
		\label{Fig. DS control}
		\vspace{-1.5em}
	\end{figure}

	\vspace{-1em}
	\subsection{Literature Review and Gap Identifications}
		The integration of heterogeneous subgrids into hybrid AC/DC/DS MGs enables effective utilization of diverse RESs but comes with increasing system complexity, particularly in terms of static and transient power management. Static power management refers to global power sharing (GPS) under steady-state operating conditions. The core objective of GPS is to ensure all generation units in the hybrid system supply power to load demand proportionally to their capacities \cite{LiFeng}. In contrast, transient power management intends to achieve that DERs with higher inertia buffer more of power demand in transient state, while those DERs with lower power density release power in a much slower way during the dynamic processes \cite{Loh}. 

	In light of static power management, for AC systems, each generation units are governed by frequency–power ($f–P$) droop. As there is only one frequency quantity in steady state, the locally measured frequency deviation by each generation can act as the coordination signal and drive the generator units to supply power commensurate with their capacity \cite{Lee}. In this way, active power sharing in AC systems can be ensured. In contrast, DC systems lack a frequency state and therefore employ voltage–power ($V–P$) droop, in which bus voltage deviations help to imply the changes in load demands \cite{ChenJianbo}. When it comes to the situation where AC and DC subgrids are bridged to form up a hybrid AC/DC system, as in reported \cite{LohPoh}, the per-unit (p.u.) values of AC frequency and DC voltage are utilized to indicate each subgrid's reserve margin. The p.u. values are then used to generate the power references for ILCs and coordinate inter-subgrid power exchanges. In this way, the global power sharing across the holistic hybrid MG can be achieved. In \cite{Jayan}, a p.u. power-consistency term is incorporated into the model predictive control for ILCs, which equalizes the contribution levels across subgrids and ensures a global power sharing. In \cite{LinPengfeng2019TSG}, a more advanced configuration involves connecting multiple ILCs in parallel to form an ILC community in the hybrid AC/DC MG. This configuration allows that each ILC in the community bears the power interaction between AC and DC subgrids in proportional to their capacity, thus to effectively avoid any possible power congestion between subgrids. Although the aforementioned static global power sharing control strategies guarantee proportional allocation with respect to subgrid capacities, it ineluctably leaves the transient behaviors of subgrids less attended. 
	
	As to dynamic power regulation, note that the extensive integration of low-inertia power electronic converters in hybrid AC/DC/DS MGs remarkably attenuates the system's overall inertia. This fact entails the hybrid MG system being sensitive to rapid load disturbances, which would result in steep frequency/voltage variations during transient periods \cite{C.L}. Protective relays may then malfunction upon detecting high rate of change of frequency/voltage (RoCoF / RoCoV) incidents and precipitate the disconnection of RESs \cite{HongQiteng, LiChang, IEC61727 } . To tackle this issue, refs. \cite{LinPengfeng2025TIE} and \cite{ZhangZhe2023} propose to modify the power control loop of ILC to enable transient power interactions between subgrids. The control methods fully mobilize global inertia resources to cope with load and RES power fluctuations. Ref. \cite{WangJie} introduces an ILC power control loop design which objective is to equalize the p.u. values of frequency / voltage rate of change in subgrids during system transition. By doing so, bi-lateral inertia support among AC and DC subgrids can be realized, and then transient performance of low-inertia subgrid can be considerably improved. However, it should be mentioned that all above research \cite{LinPengfeng2025TIE, ZhangZhe2023, WangJie} provincially focus on dynamic inertia transfer, wherein a strong assumption is made that the normalized values of maximum allowable AC frequency ($f$) and DC / DS bus voltage ($V_{dc}$ or $V_{ds}$) should be the same. This assumption, unfortunately, does not align with the condition of GPS in steady state, which means the dynamic power regulation and static GPS cannot be concurrently realized.

	\vspace{-1em}
	
	%\subsection{[C 0.1][C 1.1][C 2.1][C 3.1][C 3.2] Motivations and Contributions}
	\subsection{Motivations and Contributions}
	{In fact, dynamic inertia sharing and static GPS take place at different timescales: the former happens in transient state while the latter in steady state. No contradictions between the two objectives are observed and there should be a way to smoothly string them. In the event of load step-up or step-down, transient inertia transfer is expected to occur first and subsequently, GPS can be attained at the end of load changes. Moreover, as demonstrated in \cite{LinPengfeng2019TSG_2}, the deviations of $f$, $V_{dc}$ or $V_{ds}$ will not only impair frequency- and voltage-sensitive loads but may also induce system-wide instability. The synchronization with the utility grid also requires $f_{ac}$, $V_{dc}$ and $V_{ds}$ being driven to their nominal values as close as possible \cite{LinPengfeng2019TSG_2}. Responding to the needs of stringing dynamic and static power control objectives as well as mitigating the deviations of $f$, $V_{dc}$, $V_{ds}$ in steady sate, a novel full-timescale (FTS) power scheduling strategy for hybrid AC/DC/DS MGs is proposed.

	In this FTS power scheduling strategy, an FTS dynamic concatenator is proposed to chain up the objectives of dynamic and static power regulations at distinctive timescales. A relative loading index for retrieving the hidden loading conditions in each subgrid is also proposed. Then, autonomous frequency frequency/voltage restoration controls for AC, DC, and DS subgrids are actively consolidated into the respective control loops of the main power sources in individual subgrids, without compromising FTS power scheduling at all. The contributions of this paper are summarized as follows.}
	
	\begin{itemize}
	
		{\item An FTS dynamic concatenator is proposed to reconfigure the ILC’s power control loops accommodating transient and static behaviors of all subgrids. The dynamic concatenator is arguably reported in this paper for the first time. It makes a breakthrough that enables seamless concatenation of global transient inertia transfer to static power sharing, thereby achieving the intended power scheduling across the hybrid MG in full timescale.}
		                                                      
		% \item A steady-state autonomous frequency/voltage restoration control strategy is proposed to compensate for the steady-state deviations of $f$, $V_{dc}$, and $V_{ds}$ introduced by droop controls. DC and DS bus voltages as well as AC frequency (as seen in Fig. \ref{Fig. AC/DC/DS MG}) are autonomously restored to their nominal values. Thus, the risk of devices and sensitive loads dysfunction induced by large frequency and voltage deviations can be mitigated. A novel RLI for each subgrid is proposed to extract the hidden loading situations even if there are no obvious deviations on  $f$, $V_{dc}$, and $V_{ds}$ can be observed.
		
		\item A unified modeling approach for transient and static power management is proposed. In the approach, AC, DC, and DS subgrids are represented by Thévenin equivalent circuit models (ECMs). Three models are subsequently consolidated to a global equivalent circuit model (GECM) to characterize the entire hybrid AC/DC/DS MG. The proposed GECM provides insights to uniformly quantify the transient inertia transfer and static GPS in merely one dynamic model, with the FTS power scheduling and autonomous frequency/voltage restorations. This maneuver definitely lowers down the barriers for engineering practitioners to better dispatch inertia and power generations throughout the hybrid system.
	\end{itemize}
	
	% \vspace{-0.5em}
	
	The remainder of this paper is structured as follows: Section II and III demonstrate the configuration of hybrid AC/DC/DS MGs and the proposed FTS control strategies. Section IV presents the equivalent circuit models that elucidates the unified characterization of transient and static power management, with frequency/voltage autonomous restoration involved in each subgrid's control strategy. Sections V verifies the feasibility and effectiveness of the proposed FTS power scheduling method by experiments. Finally, Section VI concludes this paper.

	\section{Configuration of Hybrid AC/DC/DS MG}
	\subsection{Hybrid AC/DC/DS MG Configuration}
	Fig. \ref{Fig. AC/DC/DS MG} illustrates the configuration of a hybrid AC/DC/DS MG. DC-based RESs and loads like PVs and DC charging stations are integrated into DC subgrid, whereas AC-based sources including microturbines and wind turbines are connected to AC subgrid.
	 The architecture that directly matches DC-output RESs with DC loads and AC-output RESs with AC loads effectively reduces the intermediate energy conversion stages and improves system efficiency \cite{WangPeng}. A two-stage ILC consisting of a DC/DC converter ($\rm{ILC_{1}}$) and a DC/AC converter ($\rm{ILC_{2}}$) is deployed between AC and DC subgrids to manage the power transfer across subgrids. A DS subgrid, composed of a hybrid energy storage system (HESS), is integrated into the bus between $\rm{ILC_{1}}$ and $\rm{ILC_{2}}$ to buffer power imbalances incurred in AC and DC subgrids.

	\subsection{Virtual Inertia Control of AC, DC, and DS Subgrids}
     {Fig. \ref{Fig. ac and dc control} shows the virtual inertia control of AC and DC subgrids wherein the dynamic behavior of synchronous generators (SGs) is embedded into converter's control loops\cite{ZhangZhe2023}. In this control, $T$($s$)=$1/({T_G}s + 1)$, refers to the transfer function of speed governor, while $Y(s)$=$({F_{HP}}{T_{RH}}s + 1)/[({T_{CH}}s + 1)({T_{RH}}s + 1)]$, represents the transfer functions of turbine in SGs. $F_{HP}$, $T_{CH}$ and $T_{RH}$ are time constants. In Fig. \ref{Fig. ac and dc control}(a), the inertia and damping block constitute the swing equation, which characterizes the dynamics of frequency regulation of SGs. Its dynamic model in per unit (p.u.) domain\cite{Kundur} is provided as follows, where the prefix ‘$\Delta$’, subscript ‘$\rm{max}$’, and subscript ‘*’ are employed to differentiate  the change, maximum, and p.u. value of electrical quantities.}
	\begin{equation}
		\label{eq. virtual inertia of ac}
		P_{mac}^* - P_{oac}^* = 2{H_{ac}}\frac{{{\rm{d}}\Delta {f^*}}}{{{\rm{d}}t}} + {D_{ac}}\Delta {f^*}
	\end{equation}
	where $H_{ac}$ represents the inertia coefficient, while $D_{ac}$ signifies the damping coefficient of AC subgrid. $P_{{\emph{mac}}}^*$ represents the p.u. power references generated by the virtual inertia controller, and $P_{{\emph{oac}}}^*$ represents the output power. $\Delta {f^*}$ means p.u. frequency change. As shown in (\ref{eq. virtual inertia of ac}), a high inertia coefficient moderates RoCoF ( d$\Delta {f^*}$/d$t$) under abrupt load fluctuations.
	%, while a high damping component mitigates the extent of frequency change $\Delta {f^*}$. %In AC subgrid, the inverter operates in grid-forming mode, where $\Delta {f^*}$ is fed into the inner loop and determines the AC subgrid frequency.
	
	Similarly, virtual inertia control can also be extended to DC systems to enhance the transient performance of DC bus voltage \cite{ZhangZhe2023}. Fig. \ref{Fig. ac and dc control}(b) shows the virtual inertia-based control block diagram of DC subgrid. The dynamic model of DC bus voltage is as follows,
	\begin{equation}
		\label{eq. virtual inertia of dc}
		P_{mdc}^* - P_{odc}^* = 2{H_{dc}}\frac{{{\rm{d}}\Delta V_{_{dc}}^*}}{{{\rm{d}}t}} + {D_{dc}}\Delta V_{_{dc}}^*
	\end{equation}
	where $H_{dc}$ represents the inertia coefficient, while $D_{dc}$ signifies the damping coefficient of DC subgrid.  $P_{{\emph{mdc}}}^*$ represents the p.u. power references generated by the virtual inertia controller, and $P_{{\emph{odc}}}^*$ represents the output power. $\Delta V_{_{dc}}^*$ is p.u. DC bus voltage change. As virtual inertia control in both AC and DC subgrids has been extensively reported, more detailed explanations can be found in \cite{LinPengfeng2025TIE, ZhangZhe2023, WangJie}.

	{Within the DS subgrid, energy storages ($\rm{ES}$s) are divide into the cluster of low ramp rate $\rm{ES}$ ($\rm{ES_L}$) and the cluster of high ramp rate $\rm{ES}$ ($\rm{ES_H}$). $\rm{E{S_L}}$ and $\rm{E{S_H}}$ are coordinated such that they respond to different frequency components of load demand. Specifically, $\rm{E{S_L}}$ handles the slowly varying portion of load power through conventional $V$–$P$ droop control, as expressed below,
	\begin{equation}
		\label{eq. droop of ESL}
		V_{ds}^* = V_{ds\_ref}^* - \frac{1}{{{y_L}}}P_L^*
	\end{equation}
	where $P_{L}^*$ refers to p.u. output power of $\rm{E{S_L}}$ converter. $y_{L}$ is the droop coefficient of $\rm{E{S_L}}$ converter. $V_{ds}^*$ and $V_{ds\_\rm{max}}^*$ represent the p.u. output voltage and max output voltage of $\rm{E{S_L}}$ converter.
	
	{In contrast to $\rm{E{S_L}}$, $\rm{E{S_H}}$ exhibits superior performance in handling high-frequency load components due to its faster dynamic response. To enable $\rm{E{S_H}}$ to actively compensate for high-frequency load, an integral droop (ID) control strategy is employed, wherein the power term of the conventional $V$–$P$ droop control is substituted by the integral of power \cite{LinPengfeng2018TPE}:
	\begin{equation}
		\label{eq. droop of ESH}
		V_{ds}^* = V_{ds\_ref}^* - \frac{1}{{2{y_H}}}\int {P_H^*dt}
	\end{equation}
	where $P_{H}^*$ refers to p.u. output power of $\rm{E{S_H}}$ converter. $y_{H}$ is the ID coefficient of the $\rm{E{S_H}}$ converter.
	}
	
	It is worth noting that $\rm{E{S_H}}$ and $\rm{E{S_L}}$ are connected in parallel to the DS bus. They share the same output voltage, which equals the DS bus voltage shown in Fig. \ref{Fig. AC/DC/DS MG}. $\rm{E{S_H}}$ and $\rm{E{S_L}}$ are responsible for compensating high and low frequency components of the DS load power, respectively. As a result, the sum of their output powers should be equal to the total power demand of the DS subgrid, which is given in the following,
	\begin{equation}
		\label{eq. PH ad PL}
		{P_H} + {P_L} = {P_{ods}}
	\end{equation}
	
	Since the base values of $P_H$, $P_L$ and $P_{ods}$ are all the capacity of DS subgrid, the p.u. form of (\ref{eq. PH ad PL}) is as follows,
	\begin{equation}
		\label{eq. p.u. PH ad PL}
		P_L^* + P_H^* = P_{ods}^*
	\end{equation}
	
	According to (\ref{eq. droop of ESL})–(\ref{eq. p.u. PH ad PL}), the frequency domain expressions of $P_{L}^*$ and $P_{H}^*$ can be derived as,
	\begin{equation}
		\label{eq. High and low frequency distribution of ESL and ESH}
		\left\{ {\begin{array}{*{20}{c}}
				{P_L^* = \frac{{{y_L}/{y_H}}}{{s + {y_L}/{2y_H}}}P_{ods}^*}\\
				{P_H^* = \frac{s}{{s + {y_L}/{2y_H}}}P_{ods}^*}
		\end{array}} \right.
	\end{equation}
	where $\emph{s}$ is Laplace operator.
	
	It can be observed from (\ref{eq. High and low frequency distribution of ESL and ESH}) that the coordination of conventional $V$–$P$ droop and ID control enables $\rm{E{S_H}}$ and $\rm{E{S_L}}$ to handle the high and low frequency components of DS subgrid load demand, thereby utilizing the advantages of different types ESs in terms of energy and power density. 
	
	%Fig. \ref{Fig. DS control} shows the control block diagrams of $\rm{E{S_L}}$ and $\rm{E{S_H}}$ and their analogy with the swing equation of SGs. 
	Consolidating (\ref{eq. droop of ESL}) and (\ref{eq. droop of ESH}) gives the transfer function from ${P_{ods}^*}$ to ${\Delta V_{ds}^*}$, 
	\begin{equation}
		\label{eq. virtual inertia of ds}
		\frac{{\Delta V_{ds}^*}}{{\ P_{ods}^*}} = \frac{-1}{{{2y_H}s + {y_L}}}
	\end{equation}

	The relationship between  ${P_{ods}^*}$ and ${\Delta V_{ds}^*}$ indicates that the integration of conventional $V$–$P$ droop and ID intrinsically forms a dynamic frequency regulation pattern that is analogous to the swing equation of SGs, as illustrated in Fig. \ref{Fig. DS control}. Although the DS subgrid’s control loop is not explicitly incorporate virtual inertia or damping modules, the ID indeed emulates inertia behaviors, while the conventional $V–P$ droop introduces damping characteristics into DS subgrid.}
	
	\subsection{Autonomous Frequency/Voltage Restoration Strategy for AC, DC, and DS subgrids}
\begin{figure}[t]
		\setlength{\abovecaptionskip}{-3pt}
		\setlength{\belowcaptionskip}{-5pt}
		\centering
		\includegraphics[scale=1]{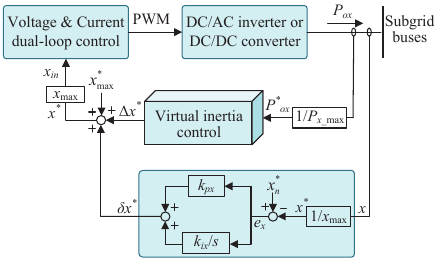}
		\caption{Autonomous frequency/voltage restoration controls for AC, DC, and DS subgrids ($x \in \left\{ f,\ V_{dc},\ V_{ds} \right\}$). The virtual inertia control can be seen in Fig. \ref{Fig. ac and dc control} and Fig. \ref{Fig. DS control}, respectively.  }
		\label{Fig. frequency voltage restoraion}
		\vspace{-1.5em}
	\end{figure}

	Due to the droop characteristics of subgrids, there are deviations in DC and DS bus voltage as well as AC frequency, with respect to their nominal values in steady-state. To avoid potential degradation or malfunction to critical equipment resulting from undesired frequency and voltage deviations \cite{YangPengcheng}, an autonomous frequency/voltage restoration strategy is presented, as seen in Fig. \ref{Fig. frequency voltage restoraion}. In this figure, $x \in \left\{ f,\ V_{dc},\ V_{ds} \right\}$, represents the measured value of frequency or bus voltage. $x_{max} \in \left\{ f_{\rm{max}},\ V_{dc\_\rm{max}},\ V_{ds\_\rm{max}} \right\}$, refers to the maximum value of $x$. $x_n^* \in \left\{ f_n^*,\ V_{dcn}^*,\ V_{dsn}^* \right\}$, refers to p.u. nominal value of $x$. $P_{ox} \in \left\{ P_{oac},\ P_{odc},\ P_{ods} \right\}$, defines the output power of each subgrid. $P_{ox}^*$ refers to the p.u. value of $P_{ox}$. $e_x$ is the error signal processed by proportional-integral (PI) controllers. $k_{px}$ and $k_{ix}$ are proportional and integral coefficients, respectively. $x_{in}$ defines the inner loop reference value. $\delta x^* \in \left\{\delta f^*, \delta V_{dc}^*, \delta V_{ds}^* \right\}$, is p.u. compensation signal for $x^*$.  As illustrated in Fig. \ref*{Fig. frequency voltage restoraion}, the PI controller gradually regulate errors to be zero by continuously generating compensation signals for virtual inertia control loops, to achieve autonomous restorations of frequency and voltage  to their nominal values.  
	% comparing the  measured frequency/voltage with their references, thereby enabling autonomous frequency and voltage restoration to their nominal values.  
	%\textcolor {black} {It is noteworthy that the proposed frequency/voltage autonomous restoration control enables  frequency and voltage to their nominal values in steady state. }
	As indicated by \cite{Luxiaonan}, PI control parameters $k_{px}$ and $k_{ix}$ can be intentionally set to small values. The PI control bandwidths are much lower than that of inertia response in each subgrid. During transient periods, the change in compensation term $\delta x^* $ can be negligible compared with the transient variation $\Delta x^* $. $\delta x^*$ has a negligible impact on the dynamic behavior of frequency and voltage. In this way,
	the intended tri-lateral inertia sharing among three subgrids would be maintained whereas no deviation of $f$, $V_{dc}$, and $V_{ds}$ in steady state would be observed.

	%Since the PI controller parameters $k_{px}$ and $k_{ix}$ are set to relatively small values, as understood from  \textcolor{red}{cite a paper from Xiaonan LU or Xia Yanghong}, its control bandwidth is  much lower than that of the inertia response. Consequently, during transient periods, the compensation term $\delta x^* $ is negligible compared with the transient variation $\Delta x^* $, thereby exerting an almost imperceptible influence on the dynamic behavior of frequency and voltage.
	%xiaonan LU low 
	
	\section{Proposed Full-Time-Scale Power Management of Hybrid AC/DC/DS MG}
	
	%\subsection{Proposed Full-Time-Scale Dynamic Concatenator for Tow-Stage ILC}
	
	As aforementioned in Section I, transient inertia sharing and static GPS happen at different timescales.These two control objectives can be carefully concatenated to achieve the intended FTS power scheduling.
	
	\subsection{Control objective I: Transient tri-lateral inertia transfer to improve the dynamic response of low-inertia subgrids}
	
	According to (\ref{eq. virtual inertia of ac}), (\ref{eq. virtual inertia of dc}), and (\ref{eq. virtual inertia of ds}), the changes of frequency and voltage ${\Delta x^*}$ can serve as indicators of the subgrid inertia . For high-inertia subgrids, ${\Delta x^*}$ changes comparatively slower during transients and exhibits a smaller dip than low-inertia subgrids. The two-stage ILC is then controlled to modulate inter-subgrid power exchange across all the subgrids so that ${\Delta x^*}$ can always be kept the same in transient state. Therefore, transient disturbances happening to the entire hybrid MG would be primarily handled by high-inertia subgrids. The low-inertia subgrids experience the mitigated frequency/voltage excursions. This fact is equivalent to the partial transfer of inertia from high-inertia subgrids to low-inertia subgrids \cite{LinPengfeng2025TIE,ZhangZhe2023}. 
	
	Based on above reasoning, control objective I of ILC should be formulated as the following,
	
	\begin{equation}
		\label{eq. equivalent frequency and voltage}
		\textbf{Control objective I:} \ \ \ \Delta {f^*} = \Delta V_{dc}^* = \Delta V_{ds}^*
	\end{equation}	
	
	Given a particular subgrid where there are multiple distributed generation units, the quantity $\Delta x^* $ is not identical among different units, which prevents the direct use of a unified $\Delta x^* $ in control implementation. According to Fig. \ref{Fig. frequency voltage restoraion}, $\Delta {x^*} = {x^*} - x_{\max }^* - \delta {x^*}$. $ x^*$ can be directly measured from the subgrid bus. $x_{\max }^*$ is a known constant. The frequency/voltage compensation term $\delta x^*$ can be obtained from a central console. As explained earlier, it's control bandwidth is tuned far slower than inertia response and it can be seen as a constant \cite{Luxiaonan}. In this sense, the control input to ILC is computed as ${x^*} - x_{\max }^* - \delta {x^*}$ which is equal to $\Delta x^*$. To facilitate theoretical derivations, the variable $\Delta x^* $ is still adopted in the subsequent analysis.

	To realize control objective I as in (\ref{eq. equivalent frequency and voltage}), the power control laws for $\rm{ILC_{1}}$ and $\rm{ILC_{2}}$ are given as,
	\begin{equation}
		\label{eq. the control of ILC with transient inertia transfer}
		\left\{ {\begin{array}{*{20}{c}}
				{{P_{1ref}} = (\Delta V_{ds}^* - \Delta V_{dc}^*){G_{\rm{PI1}}}}\\
				{{P_{2ref}} = (\Delta V_{ds}^* - \Delta {f^*}){G_{\rm{PI2}}}}
		\end{array}} \right.
	\end{equation}
	where $P_{1ref}$ and $P_{2ref}$ are power references of $\rm{ILC_{1}}$ and $\rm{ILC_{2}}$, respectively. $G_{\rm{PI1}}$ and $G_{\rm{PI2}}$ are PI controllers.

	According to (\ref{eq. the control of ILC with transient inertia transfer}), the transient power regulation objective of $\rm{ILC_{1}}$ aims to eliminate the imbalance between $\Delta V_{ds}^*$ and $\Delta V_{dc}^*$.

    When DC subgrid experiences a load disturbance, a mismatch appears between $\Delta V_{dc}^*$ and $\Delta V_{ds}^*$. According to \eqref{eq. the control of ILC with transient inertia transfer}, this mismatch generates the power reference for $\rm{ILC_{1}}$ and drives $\rm{ILC_{1}}$ to transfer power from the DS subgrid to the DC subgrid, such that the voltage deviation in DC subgrid can be alleviated and the voltage deviation of DS subgrid can be worsened. Given the high bandwidth of PI controller in the two-stage ILC, the difference between $\Delta V_{ds}^*$ and $\Delta V_{dc}^*$ can therefore be quickly regulated to approach zero with ILC1 tightly and actively regulating the power transfer between DC and DS subgrids. By doing so, part of DS subgrid inertia is shared with DC subgrid.
		
	A similar regulation occurs between the DS and AC subgrids through $\rm{ILC_{2}}$. When the AC subgrid experiences a load disturbance, a mismatch appears between $\Delta f^*$ and $\Delta V_{ds}^*$, and this mismatch generates the power reference for $\rm{ILC_{2}}$ and drives $\rm{ILC_{2}}$ to transfer power from the DS subgrid to the AC subgrid, such that the frequency deviation in AC subgrid can be alleviated and the voltage deviation of DS subgrid can be worsened. In this way, the difference between $\Delta f^*$ and $\Delta V_{ds}^*$ can be regulated to rapidly approach zero. In this way, part of DS subgrid inertia is successfully transferred to AC subgrid.

	  $\rm{ILC_{1}}$ actively redistributes part of load power to DS subgrid to equalize the transient values of $\Delta V_{ds}^*$ and $\Delta V_{dc}^*$. A similar regulation occurs between the DS and AC subgrids through $\rm{ILC_{2}}$, enabling the transient inertia transfer across the entire hybrid AC/DC/DS MG.
	
	However, transient inertia transfer focuses on the transient state power distribution among subgrids; it fails to account for steady-state power coordination. As a result, subgrids with limited capacity may be forced to supply power beyond their rated limits, whereas those with larger capacity are just lightly loaded.
	
	From Fig. \ref{Fig. ac and dc control} and Fig. \ref{Fig. DS control}, the transfer functions from p.u. output power of each subgrid to the corresponding p.u. frequency/voltage deviations can be derived. When ${s \to 0}$, these transfer functions characterizing the steady-state relationship between the output power and frequency/voltage deviations are given as,
	
	\begin{equation}
	\label{eq. steady droop of hybrid system}
	\frac{{\Delta {f^*}}}{{P_{oac}^*}} = \frac{{ - {R_{ac}}}}{{{D_{ac}}{R_{ac}} + 1}},\frac{{\Delta V_{dc}^*}}{{P_{odc}^*}} = \frac{{ - {R_{dc}}}}{{{D_{dc}}{R_{dc}} + 1}},\frac{{\Delta V_{ds}^*}}{{P_{ods}^*}} = \frac{{ - 1}}{{{y_L}}}
	\end{equation}
	
	The steady-state relations in \eqref{eq. steady droop of hybrid system} can be interpreted as the equivalent droop coefficients of each subgrid in p.u. domain. As reported in \cite{LinPengfeng2025TIE}, droop equations are normally designed in SI domain so that the all distributed sources proportionally share the holistic load power. To ensure that droop coefficients are equivalent in two domains, the parameters $R_{ac}$, $R_{dc}$, and $y_L$ in \eqref{eq. steady droop of hybrid system} should be carefully designed. The relation between the deviation of $x$ and subgrid output power ($P_{oac}$, $P_{odc}$, $P_{ods}$) in SI domain can be listed below,
	
	\begin{equation}
		\label{eq. Droop control in SI}
		\left\{ \begin{array}{l}
			\frac{{\Delta f}}{{P_{oac}^{}}}{\rm{ = }} - \frac{{{f_{\max }} - {f_{\min }}}}{{{P_{ac\_\max }}}}\\
			\frac{{\Delta V_{dc}^{}}}{{P_{odc}^{}}} =  - \frac{{{V_{dc\_\max }} - {V_{dc\_\min }}}}{{{P_{dc\_\max }}}}\\
			\frac{{\Delta V_{ds}^{}}}{{P_{ods}^{}}} =  - \frac{{{V_{ds\_\max }} - {V_{ds\_\min }}}}{{{P_{ds\_\max }}}}
		\end{array} \right.
	\end{equation}
	where $P_{ac\_\max }$, $P_{dc\_\max }$, and $P_{ds\_\max }$ stand for the maximum output power of AC, DC, and DS subgrids. Converting (\ref{eq. Droop control in SI}) into its corresponding p.u. domain gives,
	
	\begin{equation}
		\label{eq. Droop control in p.u.}
		\left\{ {\begin{array}{{l}}
				{\frac{{\Delta {f^*}}}{{P_{oac}^*}} = \frac{{\Delta f/{f_{\max }}}}{{P_{oac}^{}/{P_{ac\_\max }}}} =  -\frac{{{f_{\max }} - {f_{\min }}}}{{{P_{ac\_\max }}}}\frac{{{P_{ac\_\max }}}}{{{f_{\max }}}}}\\
				{\frac{{\Delta V_{dc}^*}}{{P_{odc}^*}} = \frac{{\Delta V_{dc}^{}/{V_{dc\_\max }}}}{{P_{odc}^{}/{P_{dc\_\max }}}}{\rm{ = }} -\frac{{{V_{dc\_\max }} - {V_{dc\_\min }}}}{{{P_{dc\_\max }}}}\frac{{{P_{dc\_\max }}}}{{{V_{dc\_\max }}}}}\\
				{ \frac{{\Delta V_{ds}^*}}{{P_{ods}^*}} =  \frac{{\Delta V_{ds}^{}/{V_{ds\_\max }}}}{{P_{ods}^{}/{P_{ds\_\max }}}} =-\frac{{{V_{ds\_\max }} - {V_{ds\_\min }}}}{{{P_{ds\_\max }}}}\frac{{{P_{ds\_\max }}}}{{{V_{ds\_\max }}}}}
		\end{array}} \right.
	\end{equation}
	
	By means of parameter matching between (\ref{eq. steady droop of hybrid system}) and (\ref{eq. Droop control in p.u.}), the following formulas hold,
	
	\begin{equation}
		\label{eq. design of droop coefficients}
		\left\{ {\begin{array}{*{20}{l}}
				{\frac{{{R_{ac}}}}{{{D_{ac}}{R_{ac}} + 1}} = \frac{{{f_{\max }} - {f_{\min }}}}{{{f_{\max }}}}}\\
				{\frac{{{R_{dc}}}}{{{D_{dc}}{R_{dc}} + 1}}{\rm{ = }}\frac{{{V_{dc\_\max }} - {V_{dc\_\min }}}}{{{V_{dc\_\max }}}}}\\
				{\frac{1}{{{y_L}}} = \frac{{{V_{ds\_\max }} - {V_{ds\_\min }}}}{{{V_{ds\_\max }}}}}
		\end{array}} \right.
	\end{equation}
	
	The expressions of $R_{ac}$, $R_{dc}$, and $y_L$ can be further derived by rearranging (\ref{eq. design of droop coefficients}), which are
	
	\begin{equation}
		\label{eq. design of droop coefficientsV2}
		\left\{ {\begin{array}{*{20}{l}}
				{{R_{ac}} = \frac{{{f_{\max }} - {f_{\min }}}}{{{f_{\max }} - {D_{ac}}({f_{\max }} - {f_{\min }})}}}\\
				{{R_{dc}}{\rm{ = }}\frac{{{V_{dc\_\max }} - {V_{dc\_\min }}}}{{{V_{dc\_\max }} - {D_{dc}}({V_{dc\_\max }} - {V_{dc\_\min }})}}}\\
				{{y_L} = \frac{{{V_{ds\_\max }}}}{{{V_{ds\_\max }} - {V_{ds\_\min }}}}}
		\end{array}} \right.
	\end{equation}
	
	Referring to \cite{Kundur} and \cite{ZhangZhe2023}, $D_{ac}$ and $D_{dc}$ are commonly configured as 1 or 2. Then $R_{ac}$ and $R_{dc}$ can also be determined by (\ref{eq. design of droop coefficientsV2}) as the maximum and minimum values of frequency and voltages are normally known in system design phase.
	
	It is worth noting that the ILC control objective stipulated in (\ref{eq. equivalent frequency and voltage}) is merely for the tri-lateral inertia sharing among AC, DC, and DS subgrids during transient process. It does not cater to the desired GPS in steady state. This fact can be validated by combining (\ref{eq. equivalent frequency and voltage}), (\ref{eq. steady droop of hybrid system}) and (\ref{eq. design of droop coefficients}). In the case that the two-stage ILC only fulfill control objective I, the ratio of output power in three subgrids in steady state can be obtained as in (\ref{eq. relationship of output power}) where $f_{\max}$, $V_{dc\_\max }$, and $V_{ds\_\max }$ are taken as base values.
	
	\begin{equation}
	\label{eq. relationship of output power}
	\begin{array}{*{20}{l}}
		{P_{oac}^*:P_{odc}^*:P_{ods}^*}\\
		\begin{array}{l}
			= \frac{{{f_{\max }}}}{{{f_{\max }} - {f_{\min }}}}:\frac{{{V_{dc\_\max }}}}{{{V_{dc\_\max }} - {V_{dc\_\min }}}}:\frac{{{V_{ds\_\max }}}}{{{V_{ds\_\max }} - {V_{ds\_\min }}}}\\
			= \frac{1}{{1 - f_{\min }^*}}:\frac{1}{{1 - V_{dc\_\min }^*}}:\frac{1}{{1 - V_{ds\_\min }^*}}
		\end{array}
	\end{array}
	\end{equation}
	
	It can be seen from (\ref{eq. relationship of output power}) that GPS can be achieved only when the permissible p.u. deviation ranges of $f$, $V_{dc}$, and $V_{ds}$ are identical. This requirement considerably hinders flexible system expansion and is too strict to be met in real applications. 
	
	\subsection{Control objective II: Static global power sharing to ensure proportional power allocation among subgrids according to their respective ratings}
	
	In steady state, the control objective of the ILC is to ensure that the output power of each subgrid is proportional to its maximum capacity, as shown in (\ref{eq. global power sharing}).  
	\begin{equation}
	\label{eq. global power sharing}
	\textbf{Control objective II:} \ \ \frac{{P_{oac}^{}}}{{P_{ac\_\max }^{}}} = \frac{{P_{odc}^{}}}{{P_{dc\_\max }^{}}} = \frac{{P_{ods}^{}}}{{P_{ds\_\max }^{}}}
	\end{equation}
	
	Owing to the droop characteristics of each subgrid, the steady-state values of $f$, $V_{dc}$, and $V_{ds}$ deviate from their maximum values $f_{\rm{max}}$, $V_{dc\_\rm{max}}$, $V_{ds\_\rm{max}}$ when steady-state autonomous frequency/voltage restoration control is not employed. These deviations, as indicated in (\ref{eq. loading condition without restoration}), directly reflect the loading conditions ($LC$s) of the corresponding subgrids. A larger $LC$ signifies that the subgrid is more heavily loaded, whereas LC will increase in the case of light load \cite{JinChi2014,ZhangZhe2021}.
	\begin{equation}
	 \label{eq. loading condition without restoration}
\left\{ \begin{array}{l}
    \vspace{1ex}
	L{C_{ac}} = \frac{{{f_{\max }} - f}}{{{f_{\max }} - {f_{\min }}}}\\
	\vspace{1ex}
	L{C_{dc}} = \frac{{{V_{dc\_\max }} - {V_{dc}}}}{{{V_{dc\_\max }} - {V_{dc\_\min }}}}\\
	\vspace{1ex}
	L{C_{ds}} = \frac{{{V_{ds\_\max }} - {V_{ds}}}}{{{V_{ds\_\max }} - {V_{ds\_\min }}}}
\end{array} \right.
	\end{equation}
	where $L{C_{ac}}$, $L{C_{dc}}$, and $L{C_{ds}}$ refer to $LC$s of AC, DC, and DS subgrids without steady-state recovery.
	
	However, due to the existence of  frequency/voltage restoration control strategy, the steady-state values of $f$, $V_{dc}$, and $V_{ds}$ are strictly clamped at their nominal values. The consequent problem is that the loading situations of subgrids are no longer accessible. To tackle this issue, a relative loading index ($RLI$) inherited from \cite{LinPengfeng2019TSG_2} is presented to extract the hidden loading information after $f$, $V_{dc}$, and $V_{ds}$ recovery. $RLI$s are defined as follows,
	
	\begin{equation}
	\label{eq. loading condition with restoration}
	\left\{ \begin{array}{l}
	\vspace{1ex}
	RL{I_{ac}} = \frac{{{f_{\max }} - f + \delta f}}{{{f_{\max }} - {f_{\min }}}}\\
	\vspace{1ex}
	RL{I_{dc}} = \frac{{{V_{dc\_\max }} - {V_{dc}} + \delta {V_{dc}}}}{{{V_{dc\_\max }} - {V_{dc\_\min }}}}\\
	\vspace{1ex}
	RL{I_{ds}} = \frac{{{V_{ds\_\max }} - {V_{ds}} + \delta {V_{ds}}}}{{{V_{ds\_\max }} - {V_{ds\_\min }}}}
	\end{array} \right.
	\end{equation}
	where $RL{I_{ac}}$, $RL{I_{dc}}$, and $RL{I_{ds}}$ refer to $RIL$s of AC, DC, and DS subgrids with steady-state recovery. The main idea of $RLI$ is to obtain the compensation term $\delta x$ by the communication link and remove it's influence from the numerator of (\ref{eq. loading condition without restoration}).

	To realize control objective II, similar to (\ref{eq. the control of ILC with transient inertia transfer}), the power loop control laws of $\rm{ILC_{1}}$ and $\rm{ILC_{2}}$ are given below,
	\begin{equation}
	\label{eq. the control of ILC with GPS}
	\left\{ {\begin{array}{*{20}{c}}
	{{P^{*}_{1ref}} = (RL{I_{dc}} - RL{I_{ds}}){G_{{\rm{PI1}}}}}\\
	{{P^{*}_{2ref}} = (RL{I_{ac}} - RL{I_{ds}}){G_{{\rm{PI2}}}}}
	\end{array}} \right.
	\end{equation}
	
	Retrospecting Fig. \ref{Fig. frequency voltage restoraion}, the inertia response with frequency or voltage compensation term in each subgrid can be formulated as
	\begin{equation}
		\label{eq. external characteristics}
		\left\{ {\begin{array}{*{20}{c}}
				{f = {f_{\max }} + \Delta f + \delta f}\\
				{{V_{dc}} = {V_{dc\_\max }} + \Delta {V_{dc}} + \delta {V_{dc}}}\\
				{{V_{ds}} = {V_{ds\_\max }} + \Delta {V_{ds}} + \delta {V_{ds}}}
		\end{array}} \right.
	\end{equation}
	
	Substituting (\ref{eq. external characteristics}) into (\ref{eq. the control of ILC with GPS}), the control law of GPS can be reformed as follows,
	\begin{equation}
		\label{eq. the control of ILC with GPSV2}
		\left\{ \begin{array}{l}
			\!\!P^{*}_{1ref} \!=\! [(\frac{{\Delta {V_{ds}}}}{{{V_{ds\_\max }} -\! {V_{ds\_\min }}}}) \!- \!(\frac{{\Delta {V_{dc}}}}{{{V_{dc\_\max }} - {V_{dc\_\min }}}})]{G_{\rm{PI}1}}\\[2mm]
			\!\!P^{*}_{2ref} \!=\! [(\frac{{\Delta {V_{ds}}}}{{{V_{ds\_\max }} - \!{V_{ds\_\min }}}}) \!-\! (\frac{{\Delta f}}{{{f_{\max }} - {f_{\min }}}})]{G_{\rm{PI}2}}
		\end{array} \right.
	\end{equation}
	
	Equation (\ref{eq. the control of ILC with GPSV2}) helps to explicitly unveil the relation between the changes in $f$, $V_{dc}$, $V_{ds}$ and power references for ILCs. This will benefit the design of aforementioned FTS dynamic concatenator and be explained in subsequent context.
	
  	{\color{black} 
   \subsection{Proposed full-time-scale dynamic concatenator to string up objectives I and II}
	
	As mentioned in Introduction, the two control objectives are realized at different timescales: the former happens in transient state while the latter in steady state. There is no contradictions between them, which makes room for identifying a controller to accommodate them. However, existing studies typically address either transient inertia transfer or static global power sharing individually. Inertia transfer control alone unnecessarily fails GPS in steady state, whereas GPS related control strategy refrain from touching dynamic management during system transition. By scrutinizing (\ref{eq. the control of ILC with transient inertia transfer}) and (\ref{eq. the control of ILC with GPSV2}), if the control objective of ILC follows (\ref{eq. the control of ILC with transient inertia transfer}) in transient stage and naturally degenerates to (\ref{eq. the control of ILC with GPSV2}) as the system reaches steady state, tri-lateral inertia sharing and steady-state GPS can be smoothly realized without switching control laws. Based on the above reasoning, this paper proposes the FTS dynamic concatenator ${T_x}(s)$ as below,
	
	\begin{equation}
		\label{eq. FTS}
		{T_x}(s) = \frac{{s + {\omega _x}}}{{s + {\omega _0}}}
	\end{equation}
	where ${\omega _0}$ is the cutoff frequency of FTS dynamic concatenator. ${\omega _x}$ is defined as ${\omega _{ac}}$, ${\omega _{dc}}$, and ${\omega _{ds}}$ in AC, DC, and DS subgrids.

	 The specific design of  ${\omega _x}$ is given as follows,
	\begin{equation}
		\label{eq. omega}
		\left\{ {\begin{array}{*{20}{c}}
				\vspace{1ex}
				{{\omega _{ac}} = \frac{{{\omega _0}{f_{\max }}}}{{{f_{\max }} - {f_{\min }}}}}\\
				\vspace{1ex}
				{{\omega _{dc}} = \frac{{{\omega _0}{V_{dc\_\max }}}}{{{V_{dc\_\max }} - {V_{dc\_\min }}}}}\\
				{{\omega _{ds}} = \frac{{{\omega _0}{V_{ds\_\max }}}}{{{V_{ds\_\max }} - {V_{ds\_\min }}}}}
		\end{array}} \right.
	\end{equation}
	
	\begin{figure}[t]
		\setlength{\abovecaptionskip}{-3pt}
		\setlength{\belowcaptionskip}{-5pt}
		\centering
		\includegraphics[scale=1]{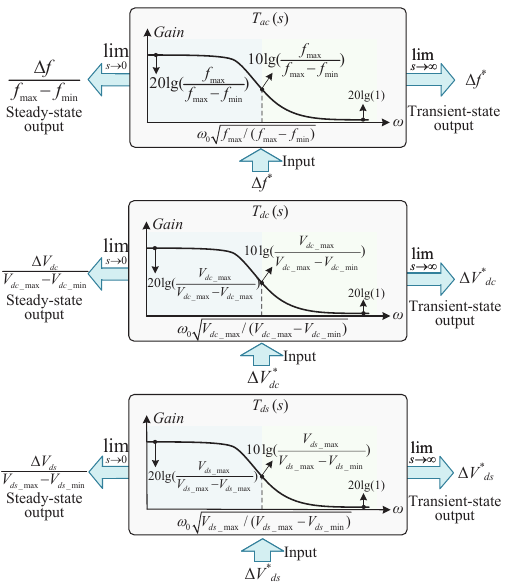}
		\caption{Schematic diagram of transient and steady-state gain of FTS dynamic concatenator, which is positioned ahead of the power PI controller within the ILC. So the power command generated by the PI controller follows (\ref{eq. the control of ILC with transient inertia transfer}) in transient stage and converges to (\ref{eq. the control of ILC with GPSV2}) in steady stage . }
		\label{Fig. Schematic diagram of FTS dynamic concatenator}
		\vspace{-1.7em} 
	\end{figure}

	\begin{figure}[t]
		\setlength{\abovecaptionskip}{-3pt}
		\setlength{\belowcaptionskip}{-5pt}
		\centering
		\includegraphics[scale=1]{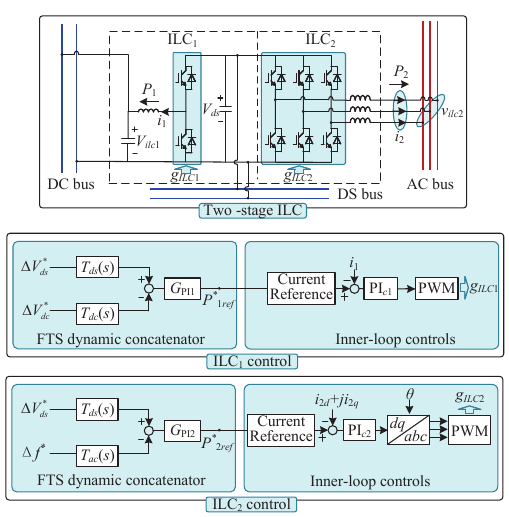}
		\caption{Control block diagram of ILC with FTS dynamic concatenator.}
		\label{Fig. ILC}
		\vspace{-1.7em}
	\end{figure}
	
	Fig. \ref{Fig. Schematic diagram of FTS dynamic concatenator} illustrates the transient and static response characteristics of FTS dynamic concatenator. The p.u. variations of frequency or voltage ($\Delta f^*, \Delta V_{dc}^*$ or $\Delta V_{ds}^* $) are fed into the FTS dynamic concatenator, of which the output values differs between the transient and steady-state stages. In the transient state (as $s$ approaches infinity), the gain of the FTS dynamic concatenator is around unity. Therefore, its outputs are $\Delta f^*, \Delta V_{dc}^*, \Delta V_{ds}^* $. While in the steady state (as $s$ approaches zero), the gain of the FTS dynamic are  $\Delta f$/($f_{\rm{max}}$-$f_{\rm{min}}$), $\Delta V_{dc}$/($V_{dc\_\rm{max}}$-$V_{dc\_\rm{min}}$), and $\Delta V_{ds}$/($V_{ds\_\rm{max}}$-$V_{ds\_\rm{mmin}}$). Based on the analyses from Fig. \ref{Fig. Schematic diagram of FTS dynamic concatenator}, the ILC control architecture depicted in Fig. \ref{Fig. ILC} is formulated. It can be observed that FTS dynamic concatenator is positioned ahead of the PI controller within the ILC, so the power command generated by the PI controller follows (\ref{eq. the control of ILC with transient inertia transfer}) in transient stage and converges to (\ref{eq. the control of ILC with GPSV2}) in steady stage. This effectively achieves the seamless transition between transient inertia transfer and static GPS, thus to enhance both transient performance and subgrid's capacity utilization efficiency.}

	%achieves the transition between transient inertia transfer and steady-state power sharing. In the transient phase, it prioritizes the inertia-supporting response by reinforcing power exchange among subgrids to suppress rapid frequency and voltage excursions. As the system approaches steady-state equilibrium, the concatenator gradually reconfigures the dynamic characteristic of ILC to proportional power allocation according to subgrid capacities. It effectively achieves seamless full-time-scale power coordination of the hybrid AC/DC/DS MG, enhancing both transient performance and steady-state capacity utilization efficiency.

	\subsection{Parameter design for ${\omega _0}$ in FTS dynamic concatenator}
	
\begin{figure}[t]
		\setlength{\abovecaptionskip}{-3pt}
		\setlength{\belowcaptionskip}{-5pt}
		\centering
		\includegraphics[scale=1]{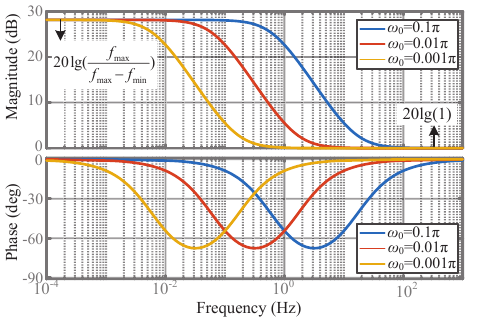}
		\caption{Bode diagrams of $T_{ac}(s)$ with different cutoff frequencies ${\omega _0}$. }
		\label{Fig. filter bode}
		\vspace{-1.5em}
	\end{figure}

	Fig. \ref*{Fig. filter bode} depicts Bode diagrams of $T_{ac}(s)$ under varying cutoff frequencies ${\omega _0}$. The decrease in ${\omega _0}$ extends the high-frequency band of the concatenator, which facilitates the transient inertia transfer. Therefore, ${\omega _0}$ should be chosen as small as possible, and there is upper limit over it. However, ${\omega _0}$ cannot be reduced indefinitely to an infinitely small value owing to numerical constraints of a digital signal processor (DSP). For the 32-bit floating-point DSP used in this work (TMS320F28335), it comprises one sign bit, eight exponent bits, and twenty-three bits forming the fraction \cite{TMS32_28335}. The resulting minimum relative separation between adjacent numbers is as follow,
	\begin{equation}
		\label{eq. varepsilon mach}
		{\varepsilon _{mach}} =  {2^{ - 23}}
	\end{equation}
	
	However, in real-time DSP computations, rounding and quantization errors accumulate through iterative updates of control states and interrupt latency. To ensure numerical stability under such conditions, an empirical safety factor $M$ is introduced and define an effective numerical resolution $\varepsilon$.
		\begin{equation}
		\label{eq. varepsilon}
	\varepsilon  = M{\varepsilon _{mach}}
	\end{equation}
	
	In particular, when FTS concatenator is discretized with a sampling period $T_{s}$, the pole of it, $s =-{\omega _0}$, is mapped into the $z$ domain as follow,
	\begin{equation}
		\label{eq. zp1}
		{z_p} = {e^{ - {\omega _0}{T_s}}}
	\end{equation}
	
	For sufficiently small ${\omega _0}$, the discrete pole approaches the unity circle, i.e.,
	\begin{equation}
		\label{eq. zp2}
		{z_p} \approx 1 - {\omega _0}{T_s}
	\end{equation}

	In finite-word-length DSP implementations, if ${\omega _0}{T_s} \le \varepsilon $, the pole is indistinguishable from unity and severe round-off errors or numerical instability may arise. Therefore, the following inequality must be satisfied by,
	\begin{equation}
		\label{eq. Omega's upper and lower limits}
		{\omega _0}{T_s} \ge \varepsilon  \Rightarrow {\omega _0} \ge \frac{\varepsilon }{{{T_s}}}
	\end{equation}
	
	For the DSP employed in this work, the sampling frequency is set to $f_s$=20 kHz, corresponding to a sampling period of $T_s$=50 $\mu$s. An empirical safety factor $M$ is set to 1.3. Hence, the minimum value of ${\omega _0}$ is $9.87 \times {10^{ - 4}}\pi$, and ${\omega _0}$ in this paper is specified as $1 \times {10^{ - 3}}\pi$. This ensures the concatenator maintains its high-frequency inertia transfer while avoiding numerical issues in the DSP implementation.

	%For the DSP employed in this work (TMS320F28335), the sampling frequency is set to $f_s$=20 kHz, corresponding to a sampling period of $T_s$=50 $\mu$s. Considering the conservative numerical resolution of a 32-bit floating-point DSP $\varepsilon=1{0^{ - 7}}$ \cite{TMS32_28335}, the lower bound of ${\omega _0}$ is set to $6.4 \times {10^{ - 4}}\pi$. This ensures the concatenator maintains its high-frequency inertia transfer while avoiding numerical degradation in DSP implementation.

	%{\color{red} [C 0.1][C 0.2][C 1.1][C 2.1] Remark: The contributions of this paper undergo beyond the simple secondary restorations of frequency or voltage as reported \cite{Guerrero2010}. As mentioned in Introduction, this paper contributes to operate HESS as a synchronous generate that can uniformly provide inertia and power coordination flexibility to the hybrid MG system. $\rm{E{S_L}}$ and $\rm{E{S_H}}$ can dynamic share the power drained from HESS. Then, a novel FTS dynamic concatenator is proposed to smoothly string up the control objectives of transient tri-lateral inertia sharing and GPS in full timescale. Moreover, RLI definitions for each subgrid are presented such that they can be integrated to the proposed FTS control strategy, without impairing the realization of inertia transfer and GPS at different timescales. Last but not least, a GECM is proposed to quantitatively characterize the transient and static power dynamics of the hybrid system. GECM helps MG system operators to better dispatch subgrid inertia and the generation of each power source.}
	
	\section{Proposed Global Equivalent Circuit Model of Hybrid AC/DC/DS MG}
	
	\subsection{Proposed Global Equivalent Circuit Model}
	The transient and static performances of AC subgrid frequency, DC and DS subgrid bus voltages have been investigated in previous sections. Section III characterizes the dynamic behavior of each subgrid by establishing its corresponding equivalent circuit model (ECM). The FTS dynamic concatenators given by (\ref{eq. FTS}) are also modeled as equivalent circuit elements. Furthermore, a global ECM (GECM) of the hybrid AC/DC/DS MG is constructed by used concatenator based elements to interconnect subgrid ECMs. The proposed GECM provides insights to uniformly quantify the transient tri-lateral inertia sharing and static GPS in merely one concise model where FTS power scheduling and autonomous frequency/voltage restorations are all accounted.
	
	To analysze the response characteristics of different subgrids, expressions in (\ref{eq. external characteristics}) are transformed into p.u. forms as described in (\ref{eq. p.u. external characteristics}), which eliminates unit discrepancies arising from different physical quantities of voltage and frequency.
	\begin{equation}
		\label{eq. p.u. external characteristics}
		{{{x^*}}}\!=\!\frac{{{x_{\max }}}}{{{x_{\max }}}}\! + \!\frac{{\Delta x}}{{{x_{\max }}}} \!+ \!\frac{{\delta x}}{{{x_{\max }}}} \Rightarrow {x^*} \!= \!1 \!+ \!\Delta {x^*} \!+ \!\delta {x^*}
	\end{equation}
	
	The corresponding frequency domain expression of (\ref{eq. p.u. external characteristics}) is,
	
		\begin{equation}
		\label{eq. p.u. external characteristics_frequency domain}
		{{{x^*}(s)}}\!= \!1/s\!+ \!\Delta {x^*(s)} \!+ \!\delta {x^*(s)}
	   \end{equation}
	
	According to the control block diagrams of subgrids shown in Fig. \ref{Fig. ac and dc control}  and Fig. \ref{Fig. DS control}, the transfer function from  p.u. output power $P_{ox}^*$ to p.u frequency/voltage $x^*$ are as follows,
	\begin{equation}
		\label{eq. equivalent impedance}
\left\{ {\begin{array}{*{20}{c}}
		{\frac{{\Delta {f^*}(s)}}{{P_{oac}^*(s)}} = \frac{{ - {R_{ac}}}}{{(2{H_{ac}}s + {D_{ac}}){R_{ac}} + T(s)Y(s)}}}\\
		{\frac{{\Delta V_{dc}^*(s)}}{{P_{odc}^*(s)}} = \frac{{ - {R_{dc}}}}{{(2{H_{dc}}s + {D_{dc}}){R_{dc}} + T(s)Y(s)}}}\\
		{\frac{{\Delta V_{ds}^*(s)}}{{P_{ods}^*(s)}} = \frac{{ - 1}}{{2{y_H}s + {y_L}}}}
\end{array}} \right.
	\end{equation}
	
	In (\ref{eq. equivalent impedance}), the power base of each subgrid is initially defined as its own maximum output power, leading to the p.u. output power $P_{oac}^*$, $P_{odc}^*$, and $P_{ods}^*$. Although this is convenient for describing local subgrid behavior, it complicates the analysis of power transfer among interconnected subgrids. Therefore, for subsequent system-level modeling, the power bases in (\ref{eq. equivalent impedance}) are unified and replaced by the total maximum capacity $P_{G\max}$ of the hybrid AC/DC/DS MG.

	\begin{equation}
	\label{eq. equivalent impedanceS2}
\left\{ \begin{array}{l}
	\frac{{\Delta {f^*}(s)}}{{P_{oac\_G}^*(s)}} = \frac{{ - {R_{ac}}}}{{\frac{{{P_{ac\_\max }}}}{{{P_{G\max }}}}(2{H_{ac}}s + {D_{ac}}){R_{ac}} + \frac{{{P_{ac\_\max }}}}{{{P_{G\max }}}}T(s)Y(s)}} =  - {Z_{ac}}(s)\\
	\frac{{\Delta V_{dc}^*(s)}}{{P_{odc\_G}^*(s)}} = \frac{{ - {R_{dc}}}}{{\frac{{{P_{dc\_\max }}}}{{{P_{G\max }}}}(2{H_{dc}}s + {D_{dc}}){R_{dc}} + \frac{{{P_{dc\_\max }}}}{{{P_{G\max }}}}T(s)Y(s)}} =  - {Z_{dc}}(s)\\
	\frac{{\Delta V_{ds}^*(s)}}{{P_{ods\_G}^*(s)}} = \frac{{ - 1}}{{\frac{{{P_{dc\_\max }}}}{{{P_{G\max }}}}(2{y_H}s + {y_L})}} =  - {Z_{ds}}(s)\\
	P_{oac\_G}^{}(s){\rm{ = }}\frac{{P_{oac}^*(s){P_{ac\_\max }}}}{{{P_{G\max }}}},\\
	P_{odc\_G}^*(s){\rm{ = }}\frac{{P_{odc}^*(s){P_{dc\_\max }}}}{{{P_{G\max }}}},\\
	P_{ods\_G}^*(s){\rm{ = }}\frac{{P_{odc}^*(s){P_{dc\_\max }}}}{{{P_{G\max }}}},\\
	{P_{G\max }} = {P_{ac\_\max }} + {P_{dc\_\max }} + {P_{ds\_\max }}.
\end{array} \right.
\end{equation}
where $P_{G\max}$ is the total maximum capacity of the hybrid AC/DC/DS MG.  $P_{ox\_G} \in \left\{ P_{oac\_G}^*,\ P_{odc\_G}^*,\ P_{ods\_G}^* \right\}$ are p.u. output powers of three subgrids by re-normalizing the actual output powers with respect to the power base $P_{G\max}$.

\begin{figure}[t]
		\setlength{\abovecaptionskip}{-3pt}
		\setlength{\belowcaptionskip}{-5pt}
		\centering
		\includegraphics[scale=1]{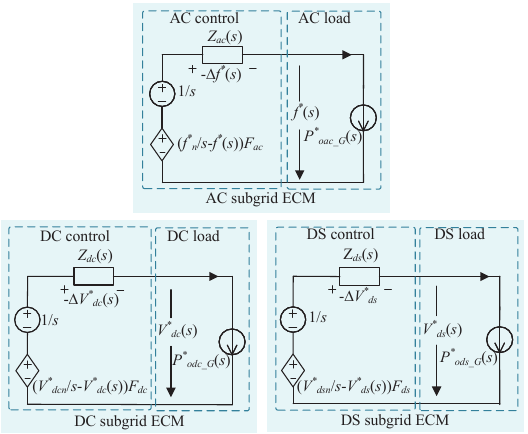}
		\caption{The equivalent circuit model of each subgrid.}
		\label{Fig. each ECM}
		\vspace{-1.5em}
	\end{figure}

	Equation (\ref{eq. equivalent impedanceS2}) can be uniformly expressed by,
	
	\begin{equation}
		\label{eq. p.u equivalent impedance}
		\frac{{\Delta {x^*}(s)}}{{P_{ox\_G}^*(s)}} =  - {Z_x}(s)
	\end{equation}
	wherein $Z_{ac}(s)$, $Z_{dc}(s)$, and $Z_{ds}(s)$ are collectively represented by $Z_x$. Substituting (\ref{eq. p.u equivalent impedance}) into (\ref{eq. p.u. external characteristics}) gives the following expression,
	\begin{equation}
		\label{eq. p.u. external characteristics in s domain}
		{x^*(s)} = (1/s + \delta {x^*(s)}) - P_{ox\_G}^*(s){Z_x(s)}
	\end{equation}
	where $(1/s + \delta {x^*(s)})$, $P_{ox\_G}^*(s)$, and $Z_x(s)$ are interpreted as open-circuit voltage, short-circuit current, and equivalent impedance, respectively. Then, equation(\ref{eq. p.u. external characteristics in s domain}) resembles the similar format of Thévenin equivalent circuit. Accordingly, ECMs depicted in Fig. \ref*{Fig. each ECM} reveals the dynamic response characteristics of each subgrid from the circuit perspective, where $\delta {x^*(s)}$ is expressed by (\ref{eq. restoration expression}) according to the autonomous frequency/voltage control diagram in Fig. \ref{Fig. frequency voltage restoraion}.
	\begin{equation}
		\label{eq. restoration expression}
		\delta {x^*(s)} = (x_n^*/s - {x^*(s)}){F_x}
	\end{equation}
	where ${F_x} = {k_{px}} + {k_{ix}}/s$, represents the transfer function of PI controller in frequency/voltage restoration controls.
	
	In the hybrid AC/DC/DS MG, the two-stage ILC equipped with the FTS dynamic concatenator functions as the channel for inter-subgrid power exchange. This means the power exchanged through the ILC couples the dynamics of all three subgrids. Consequently, constructing a global ECM that captures the overall system dynamics requires an explicit circuit modeling of the ILC. According to Fig. \ref{Fig. ILC}, the dynamic characteristics of FTS dynamic concatenator can be converted into the following equivalent impedance.
	\begin{equation}
		\label{eq. ILC impedance}
\left\{ {\begin{array}{*{20}{c}}
		{\frac{{({T_{ds}}(s)\Delta V_{ds}^*(s) - {T_{dc}}(s)\Delta V_{dc}^*(s))}}{{P_{1ref}^*{}(s)}} = \frac{s}{{{k_{tp1}}s + {k_{ti1}}}} = {Z_{ILC1}}(s)}\\
		{\frac{{({T_{ds}}(s)\Delta V_{ds}^*(s) - {T_{ac}}(s)\Delta {f^*}(s))}}{{P_{2ref}^*{}(s)}} = \frac{s}{{{k_{tp2}}s + {k_{ti2}}}} = {Z_{ILC2}}(s)}
\end{array}} \right.
	\end{equation}
	
	\begin{figure*}[t]
	\setlength{\abovecaptionskip}{-3pt}
	\setlength{\belowcaptionskip}{-5pt}
	\centering
	\includegraphics[scale=1]{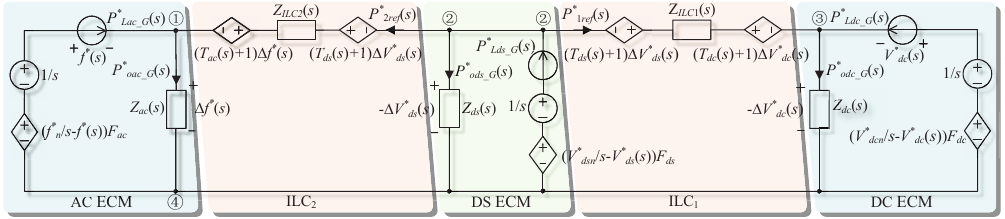}
	\caption{ Global equivalent circuit model of hybrid AC/DC/DS MG with FTS dynamic concatenator and frequency/voltage restoration control.}
	\label{Fig. global ECM}
	\vspace{-1.6em}
 	\end{figure*}
	Assuming that $Z_{\mathrm{ILC}1}(s)$ and $Z_{\mathrm{ILC}2}(s)$ in \eqref{eq. ILC impedance} denote equivalent impedances, and $P_{1ref}^*(s)$ and $P_{2ref}^*(s)$ denote the corresponding equivalent currents, the impedance voltages $\Delta f^{\ast}(s)$, $\Delta V^{\ast}_{{dc}}(s)$, and $\Delta V^{\ast}_{{ds}}(s)$ can be interconnected through the equivalent impedances $Z_{\mathrm{ILC}1}(s)$ and $Z_{\mathrm{ILC}2}(s)$. In this way, the GECM of hybrid AC/DC/DS MG shown in Fig. \ref{Fig. global ECM} is obtained. In this figure, $P_{Lx\_G}^*(s) \in \left\{ P_{Lac\_G}^*(s),\ P_{Ldc\_G}^*(s),\ P_{Lds\_G}^*(s) \right\}$ represents the local load power of the individual subgrid. All of subgrids' local load power are normalized with respect to the base value $P_{G\max}$. In the absence of ILC interconnection, the output power of each subgrid is solely determined by its local load demand. Through the interconnection by ILC, the output power of each subgrid is instead defined as the aggregation of its local load demand and the power from other subgrids.

	\subsection{Full-Time-Scale Dynamic Analysis of MG Base on GECM}
	
    Note that Fig. \ref{Fig. global ECM} visualizes the dynamic response of hybrid AC/DC/DS MG through equivalent circuit representation. GECM of the entire hybrid MG can be modeled and solved by nodal analyses, as shown below,
	\begin{equation}
		\label{eq. node voltage}
		\mathversion{bold}
		GV_{node} =  {I_P}
	\end{equation} 
	where $\boldsymbol{G}$ represents the node admittance matrix in  Fig. \ref{Fig. global ECM}. $\boldsymbol{V_{node}}$ denotes node voltage vector in which each element is the node voltage. $\boldsymbol{I_P}$ represents the vector in which each element is the current injecting into the node from current sources. The specifics of $\boldsymbol{G}$, $\boldsymbol{V_{node}}$, and $\boldsymbol{I_P}$ are given below,
		\begin{equation}	\label{eq. G, Vnode, and Ip}
\begin{array}{l}
	\boldsymbol{G} = \left[ {\begin{array}{*{20}{c}}
			{\frac{1}{{{Z_{ac}}}} - \frac{{{T_{ac}}}}{{{Z_{ILC2}}}}} &
			{\frac{{{T_{ds}}}}{{{Z_{ILC2}}}}} & 0 \\
			{\frac{{{T_{ac}}}}{{{Z_{ILC2}}}}} &
			{\frac{1}{{{Z_{ds}}}} - \frac{{{T_{ds}}}}{{{Z_{ILC1}}}} - \frac{{{T_{ds}}}}{{{Z_{ILC2}}}}} &
			{\frac{{{T_{ds}}}}{{{Z_{ILC1}}}}} \\
			0 &
			{\frac{{{T_{ds}}}}{{{Z_{ILC1}}}}} &
			{\frac{1}{{{Z_{dc}}}} - \frac{{{T_{dc}}}}{{{Z_{ILC1}}}}}
	\end{array}} \right], \\
	
	\boldsymbol{V_{node}} = \left[ {\begin{array}{*{20}{c}}
			{ - \Delta {f^*}} \\
			{ - \Delta V_{ds}^*} \\
			{ - \Delta V_{dc}^*}
	\end{array}} \right], \quad
	\boldsymbol{I_P} = \left[ {\begin{array}{*{20}{c}}
			{P_{Lac\_G}^*} \\
			{P_{Lds\_G}^*} \\
			{P_{Ldc\_G}^*}
	\end{array}} \right].
\end{array}
	\end{equation}
	
	In the following, the transient and steady-state characteristics of each subgrid’s voltage or frequency, as well as the steady-state power characteristics, are further analyzed based on $\boldsymbol{V_{node}}$ and $\boldsymbol{I_P}$ obtained from (\ref{eq. node voltage}) and (\ref{eq. G, Vnode, and Ip}).
	
	(1)  \textbf{ Characterization of transient frequency / voltage responses and the analysis of bri-lateral inertia transfer scheme, based on Initial Value Theorem}
	
	The transient inertia transfer mechanism can be revealed by comparing the inertia levels of each subgrid before and after the inertia sharing control is configured. According to (\ref{eq. virtual inertia of ac}), (\ref{eq. virtual inertia of dc}), and (\ref{eq. virtual inertia of ds}), it can be seen that $\Delta x^*(t)$ and $P_{ox}^*(t)$ are negatively correlated with the inertia coefficient. ${\Delta x^*(t)/P_{ox}^*(t)}$ attains its maximum absolute value when $t$ $\to$ ${0^+}$. Consequently, $\Delta x^*(t)/P_{ox}^*(t)|{_{t \to {0^ + }}}$ can serve as an indicator of the subgrid’s inertia level. Accordingly, before the FTS dynamic concatenator is enabled, the inertia level of each subgrid can be represented by $\Delta x^*(t)/P_{ox}^*(t)|{_{t \to {0^ + }}}$. Through Laplace transform, $\Delta x^*(t)/P_{ox}^*(t)|{_{t \to {0^ + }}}$ can be represented in the Laplace domain as,

	\begin{equation}
		\label{eq. Nx0}
		\left\{ {\begin{array}{*{20}{l}}
		{{{\left. {\frac{{\Delta {f^*}(t)}}{{P_{oac}^*(t)}}} \right|}_{t \to {0^ + }}} = {{\left. {\frac{{\Delta {f^*}(s)}}{{P_{oac}^*(s)}}} \right|}_{s \to \infty }}}\\
		{{{\left. {\frac{{\Delta V_{dc}^*(t)}}{{P_{odc}^*(t)}}} \right|}_{t \to {0^ + }}} = {{\left. {\frac{{\Delta V_{dc}^*(s)}}{{P_{odc}^*(s)}}} \right|}_{s \to \infty }}}\\
		{{{\left. {\frac{{\Delta V_{ds}^*(t)}}{{P_{ods}^*(t)}}} \right|}_{t \to {0^ + }}} = {{\left. {\frac{{\Delta V_{ds}^*(s)}}{{P_{ods}^*(s)}}} \right|}_{s \to \infty }}}
		\end{array}} \right.
	\end{equation}

To simplify the notation, let $\frac{{\Delta {f^*}(s)}}{{P_{oac}^*(s)}} = {N_{ac0}}(s),\frac{{\Delta V_{dc}^*(s)}}{{P_{odc}^*(s)}} = {N_{dc0}}(s)$, and $\frac{{\Delta V_{ds}^*(s)}}{{P_{ods}^*(s)}} = {N_{ds0}}(s)$. 

 ${N_{ac0}}(s)$, ${N_{dc0}}(s)$, and ${N_{ds0}}(s)$ can an be directly obtained from the control block diagrams of the subgrids shown in Fig. \ref{Fig. ac and dc control} and Fig. \ref{Fig. DS control}. Initial Value Theorem can then be applied to derive the inertia level of each subgrid without inertia sharing control, 
 \begin{equation}
 	\label{eq. Initial value theorem without FTS}
 	\left\{ {\begin{array}{*{20}{c}}
 			{\mathop {\lim }\limits_{s \to \infty } s{N_{ac0}}(s) = \mathop {\lim }\limits_{s \to \infty } \frac{{-s{R_{ac}}}}{{(2{H_{ac}}s + {D_{ac}}){R_{ac}} + T(s)Y(s)}} = \frac{-1}{{2{H_{ac}}}}}\\
 			{\mathop {\lim }\limits_{s \to \infty } s{N_{dc0}}(s) = \mathop {\lim }\limits_{s \to \infty } \frac{{-s{R_{dc}}}}{{(2{H_{dc}}s + {D_{dc}}){R_{dc}} + T(s)Y(s)}} = \frac{-1}{{2{H_{dc}}}}}\\
 			{\mathop {\lim }\limits_{s \to \infty } s{N_{ds0}}(s) = \mathop {\lim }\limits_{s \to \infty } \frac{-s}{{{2y_H}s + {y_L}}} = \frac{-1}{{{2y_H}}}}
 	\end{array}} \right.
 \end{equation}
 
 As shown in (\ref{eq. Initial value theorem without FTS}), applying initial value theorem to (\ref{eq. Nx0}) yields results related to the reciprocals of the inherent inertia of each subgrid. This means large inertia entails smaller deviations on frequency / voltage, and Initial Value Theorem in modern control theory can help to easily unveil inertia level in each subgrid.
 
 After the proposed FTS dynamic connector is enabled, power exchange occurs among the subgrids. The frequency or voltage of each subgrid is influenced not only by its local load but also by the loads of other subgrids. In other words, load disturbances in any region of the hybrid AC/DC/DS MG are shared among all subgrids. Accordingly, with the FTS dynamic concatenator being enabled, the inertia level of each subgrid can be represented by $\Delta x^*(t)/P_{LG}^*(t)|{_{t \to {0^ + }}}$, where $P_{LG}$ is the total load power of the hybrid AC/DC/DS MG. Through Laplace transform,  $\Delta x^*(t)/P_{LG}^*(t)|{_{t \to {0^ + }}}$ can be represented in the Laplace domain as follows,
\begin{equation}
	\label{eq. Nx1}
\left\{ {\begin{array}{*{20}{l}}
		{{{\left. {\frac{{\Delta {f^*}(t)}}{{P_{LG}^*(t)}}} \right|}_{t \to {0^ + }}} = {{\left. {\frac{{\Delta {f^*}(s)}}{{P_{LG}^*(s)}}} \right|}_{s \to \infty }}}\\
		{{{\left. {\frac{{\Delta V_{dc}^*(t)}}{{P_{LG}^*(t)}}} \right|}_{t \to {0^ + }}} = {{\left. {\frac{{\Delta V_{dc}^*(s)}}{{P_{LG}^*(s)}}} \right|}_{s \to \infty }}}\\
		{{{\left. {\frac{{\Delta V_{ds}^*(t)}}{{P_{LG}^*(t)}}} \right|}_{t \to {0^ + }}} = {{\left. {\frac{{\Delta V_{ds}^*(s)}}{{P_{LG}^*(s)}}} \right|}_{s \to \infty }}}\\
		{\begin{array}{*{20}{l}}
				{P_{LG}^{} = P_{Lac\_G}^{} + P_{Ldc\_G}^{} + P_{Lds\_G}^{}}\\
				{P_{LG}^* = \frac{{P_{LG}^{}}}{{{P_{G\_\max }}}}}
		\end{array}}
\end{array}} \right.
\end{equation}

To simplify the notation, let $\frac{{\Delta {f^*}(s)}}{{P_{LG}^*(s)}} = {N_{ac1}}(s),\frac{{\Delta V_{dc}^*(s)}}{{P_{LG}^*(s)}} = {N_{dc1}}(s)$, and $\frac{{\Delta V_{ds}^*(s)}}{{P_{LG}^*(s)}} = {N_{ds1}}(s)$.

	${N_{ac1}}(s)$, ${N_{dc1}}(s)$, and ${N_{ds1}}(s)$ can an be obtained from (\ref{eq. node voltage}). Based on this, Initial Value Theorem is applied to derive the inertia level index of each subgrid with inertia sharing control, as follows,
	\begin{equation}
		\label{eq. Nac1}
\left\{ \begin{array}{l}
	{N_{ac1}}(s) = \frac{{{R_{ac}}{R_{dc}}{T_{ac}}(s){T_{dc}}(s)}}{{[{R_{ac}}{R_{dc}}({G_{ac}}(s) + {G_{dc}}(s) + {G_{ds}}(s)) + L(s)]}}\\
	{N_{dc1}}(s) = \frac{{{R_{ac}}{R_{dc}}{T_{ac}}(s){T_{dc}}(s){P_G}(s)}}{{[{R_{ac}}{R_{dc}}({G_{ac}}(s) + {G_{dc}}(s) + {G_{ds}}(s)) + L(s)]}}\\
	{N_{ds1}}(s) = \frac{{{R_{ac}}{R_{dc}}{T_{ac}}(s){T_{dc}}(s){P_G}(s)}}{{[{R_{ac}}{R_{dc}}({G_{ac}}(s) + {G_{dc}}(s) + {G_{ds}}(s)) + L(s)]}}\\
	{G_{ac}}(s) = (2{H_{ac}}s + {D_{ac}}){T_{dc}}(s){T_{ds}}(s){P_{ac\_\max }}/{P_{G\_\max }}\\
	{G_{dc}}(s) = (2{H_{dc}}s + {D_{dc}}){T_{ac}}(s){T_{ds}}(s){P_{ac\_\max }}/{P_{G\_\max }}\\
	{G_{ds}}(s) = (2{y_H}s + {y_L}){T_{ac}}(s){T_{ds}}(s){P_{ac\_\max }}/{P_{G\_\max }}\\
	M(s) = {R_{ac}}{T_{ac}}(s){T_{ds}}(s) + {R_{dc}}{T_{dc}}(s){T_{ds}}(s)\\
	L(s) = T(s)Y(s)M(s)
\end{array} \right.
	\end{equation}
	
To ensure the implementation of inertia transfer, the PI controllers in the ILC power loop must react far more quickly than the intrinsic inertia response of each subgrid. This fast response is achieved by selecting sufficiently large proportional and integral gains $k_{tpx}$ and $k_{tix}$  of PI controllers in the ILC power loop. In this sense, (\ref{eq. Nac1}) can be further expressed as follows,
	\begin{equation}
		\label{eq. Initial value theorem with FTS}
\left\{ \begin{array}{l}
	\mathop {\lim }\limits_{s \to \infty } s{N_{ac1}}(s) = \frac{{ - 1}}{{2{H_G}}}\\
	\mathop {\lim }\limits_{s \to \infty } s{N_{dc1}}(s) = \frac{{ - 1}}{{2{H_G}}}\\
	\mathop {\lim }\limits_{s \to \infty } s{N_{ds1}}(s) = \frac{{ - 1}}{{2{H_G}}}\\
	{H_G} = \:({H_{ac}}\frac{{{P_{ac\_\max }}}}{{{P_{G\_\max }}}} + {H_{dc}}\frac{{{P_{dc\_\max }}}}{{{P_{G\_\max }}}} + {y_H}\frac{{{P_{ds\_\max }}}}{{{P_{G\_\max }}}})
\end{array} \right.
	\end{equation}
	
	The inertia transfer mechanism of hybrid AC/DC/DS MG can be exemplified by  comparing (\ref{eq. Initial value theorem without FTS}) and (\ref{eq. Initial value theorem with FTS}). $H_G$ refers to a global inertia coefficient. It is derived by weighting the inherent inertia of each subgrid, where the weighting factor for each subgrid is the ratio of its maximum power capacity to the total system capacity. It should mentioned that the expression of $H_G$ is the same as that of Center of Inertia (COI) in conventional power system \cite{Kundur}. This fact provides system operator with better understanding for the dynamic behavior of wide penetration of power electronics and they can somehow dispatch system inertia in their maturer way.
	As indicated by (\ref{eq. Initial value theorem with FTS}), inertia transfer control couples all subgrid dynamics. The inertia resources of all subgrids are pooled and collectively contribute to response to global load disturbances.  	
	
\begin{figure}[t]
		\setlength{\abovecaptionskip}{-3pt}
		\setlength{\belowcaptionskip}{-5pt}
		\centering
		\includegraphics[scale=1]{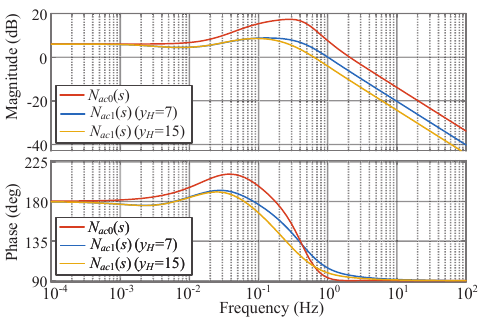}
		\caption{Bode diagrams of $N_{ac0}(s)$ and $N_{ac1}(s)$ for AC sbugrid. $N_{ac0}(s)$ represents the frequency domain expression of $\Delta x^*(t)/P_{ox}^*(t) |_{t \to {0^ + }}$ without FTS dynamic concatenator). $N_{ac1}(s)$ represents the frequency domain expression of $\Delta x^*(t)/P_{ox}^*(t) |_{t \to {0^ + }} $ with FTS dynamic concatenator).}
		\label{Fig. Bode sN}
		\vspace{-1.5em}
	\end{figure}

	Fig. \ref{Fig. Bode sN} illustrates Bode diagrams of $N_{ac0}(s)$ and $N_{ac1}(s)$ under a 60\% load perturbation in subgrids. As shown in Fig. \ref{Fig. Bode sN}, the high-frequency magnitude of $N_{ac1}(s)$ is lower than that of $N_{ac0}(s)$. This reduction indicates  the AC subgrid is supported by other subgrids with higher inertia through inertia transfer control. AC subgrid exhibits an increase in inertia level. Furthermore, when the integral gain $y_H$ is increased from 7 to 15, the magnitude of $N_{ac1}(s)$ decreases even further in high frequency band. This trend demonstrates that a larger $y_H$ effectively enhances the inertia level of the DS subgrid, and the inertia level of AC subgrid is further increased due to inertia transfer control.
	
	(2) \textbf{Characterization of steady-state frequency / voltage and power outputs of the subgrids, based on Final Value Theorem} 
	
   Upon substituting $\Delta x^*(s)$ derived from (\ref{eq. node voltage}) into (\ref{eq. p.u. external characteristics_frequency domain}), the steady-state responses of subgrid frequency and voltage can be directly analyzed via the final value theorem.
	\begin{equation}
		\label{eq. Steady-state analysis of frequency/voltage}
		\begin{aligned}
			\left. x^*(t) \right|_{t \to \infty} 
			&= \lim_{s \to 0} x^*(s)\, s \\
			&= \Bigl[\bigl(\tfrac{1}{s} + x_{n}^*(s)\bigr)/(1+F_x(s)) - P_{ox\_G}^*(s)Z_x\Bigr] s \\
			&= x_n^*
		\end{aligned}
	\end{equation}
	
	\begin{figure}[t]
		\setlength{\abovecaptionskip}{-3pt}
		\setlength{\belowcaptionskip}{-5pt}
		\centering
		\includegraphics[scale=1]{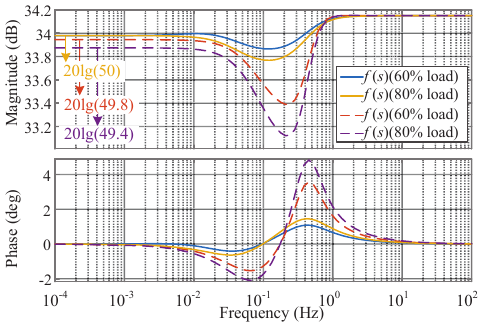}
		\caption{Bode diagrams of $f(s)$ in AC subgrid with steady-state restoration (solid lines) and $f(s)$ without steady-state restoration (dotted lines).}
		\label{Fig. Bode sf}
		\vspace{-1.5em}
	\end{figure}

 Equation (\ref{eq. Steady-state analysis of frequency/voltage}) shows that the steady-state value of $x^*(t)$ autonomously converges to $x_n^*$ due to existence of the proposed frequency/voltage restoration strategy in Section II Part C.
 
 As an example, Fig. \ref{Fig. Bode sf} shows Bode diagrams for $f(s)$ with steady-state autonomous restoration (solid lines) and $f(s)$ without steady-state autonomous restoration (dashed lines). As observed from Fig. \ref{Fig. Bode sf}, with steady-state autonomous restoration, the low-frequency magnitude of $f$ remains fixed at $20\rm{lg}(50)$ regardless of load variations. In contrast, without steady-state autonomous restoration, the low-frequency magnitude of $f(s)$ decreases progressively as the load increases.
	
	According to (\ref{eq. node voltage}), when the $Z_{ILCn}$ is ignored, the p.u. output power of each subgrid can be deduced as follows,
		\begin{equation}
		\label{eq. output power analysis of the subgrids}
\left\{ {\begin{array}{*{20}{l}}
		{P_{oac\_G}^*(s) = \frac{{(T(s)\:Y(s) + {D_{ac}}{R_{ac}} + 2{H_{ac}}{R_{ac}}s){R_{dc}}{T_{dc}}(s){T_{ds}}(s)}}{{{{G'}_{ac}}(s) + {{G'}_{dc}}(s) + {{G'}_{ds}}(s)}}}\\
		{P_{odc\_G}^*(s) = \frac{{(T(s)\:Y(s) + {D_{dc}}{R_{dc}} + 2{H_{dc}}{R_{dc}}s){R_{ac}}{T_{ac}}(s){T_{ds}}(s)P_{LG}^*(s)}}{{{{G'}_{ac}}(s) + {{G'}_{dc}}(s) + {{G'}_{ds}}(s)}}}\\
		{P_{ods\_G}^*(s) = \frac{{{R_{ac}}{R_{dc}}{T_{ac}}(s){T_{dc}}(s)({y_L} + 2{y_H}s)P_{LG}^*(s)}}{{{{G'}_{ac}}(s) + {{G'}_{dc}}(s) + {{G'}_{ds}}(s)}}}\\
		{{{G'}_{ac}}(s) = {G_{ac}}(s){R_{ac}}{R_{dc}} + {R_{ac}}T(s)\:Y(s){T_{ac}}(s){T_{ds}}(s)}\\
		{{{G'}_{dc}}(s) = {G_{dc}}(s){R_{ac}}{R_{dc}} + {R_{dc}}T(s)\:Y(s){T_{dc}}(s){T_{ds}}(s)}\\
		{{{G'}_{ds}}(s) = {G_{ds}}(s){R_{ac}}{R_{dc}}}
\end{array}} \right.
	\end{equation}
	
	By first applying the final value theorem to (\ref{eq. output power analysis of the subgrids}) and then substituting the result from  (\ref{eq. design of droop coefficientsV2}), the steady-state output power of each subgrid is obtained.
	\begin{equation}
		\label{eq. output power analysis of the subgrids S2}
\left\{ {\begin{array}{*{20}{c}}
		{{{\left. {P_{oac\_G}^*(t)} \right|}_{t \to \infty }} = \mathop {\lim }\limits_{s \to } sP_{oac\_G}^*(s) = {P_{ac\_\max }}\frac{{P_{LG}^*}}{{{P_{G\_\max }}}}}\\
		{{{\left. {P_{odc\_G}^*(t)} \right|}_{t \to \infty }} = \mathop {\lim }\limits_{s \to } sP_{odc\_G}^*(s) = {P_{dc\_\max }}\frac{{P_{LG}^*}}{{{P_{G\_\max }}}}}\\
		{{{\left. {P_{ods\_G}^*(t)} \right|}_{t \to \infty }} = \mathop {\lim }\limits_{s \to } sP_{ods\_G}^*(s) = {P_{ds\_\max }}\frac{{P_{LG}^*}}{{{P_{G\_\max }}}}}
\end{array}} \right.
	\end{equation}
	
	According to (\ref{eq. output power analysis of the subgrids S2}), the steady-state output powers of subgrids in the SI domain exhibit the following proportional relationship.
	
	\begin{equation}
		\label{eq. Output power analysis of the subgrids in steady stage}
	\begin{array}{l}
		P_{oac}^{}:P_{odc}^{}:P_{ods}^{}\\
		= P_{oac\_G}^*{P_{G\_\max }}:P_{odc\_G}^*{P_{G\_\max }}:P_{ods\_G}^*{P_{G\_\max }}\\
		= {P_{ac\_\max }}:{P_{dc\_\max }}:{P_{ds\_\max }}
	\end{array}
	\end{equation}
	
	As established by (\ref{eq. Output power analysis of the subgrids in steady stage}), the actual output power of each subgrid is proportional to its maximum capacity in steady state, and GPS is therefore obtained.
	
	\vspace{-1.0em}
	\subsection{Discussion}
	
     \textit{1) The HESS configuration}: Departing from conventional hybrid AC/DC/DS MG configurations \cite{WangPeng, JinChi2018}, an HESS is proposed to replace the traditional single energy storage unit in the DS subgrid. The HESS is composed of different $\rm{ES}$s that can be classified into $\rm{E{S_L}}$ and $\rm{E{S_H}}$. ID and traditional $V–P$ droops are applied to $\rm{E{S_H}}$ and $\rm{E{S_L}}$. $\rm{E{S_H}}$s respond to high-frequency components of load fluctuations, whereas $\rm{E{S_L}}$s compensate for low-frequency ones. This method not only makes full use of the various advantages of different $\rm{ES}$s, but also coordinates HESS as a whole to work as a DC generator. In this way, the HESS in DS subgrid can be dispatched to uniformly provide inertia supports to AC / DC subgrids and enhance the control flexibility of overall system.
	
	\textit{2) The motivation of introducing GECM}: On one hand, a hybrid AC/DC/DS microgrid may have various types of power electronic converters interacting with each other. The complex system architecture would cause difficulties to analyze system performance. On the other hand, the power converters and subgrids are normally programmed with closed-loop controllers. Knowledge of steady-state performance alone is therefore insufficient to predict microgrid dynamic behavior and interactions under disturbances. Many researchers and engineers heavily rely on simulation tools to capture microgrid’s static and dynamic performance, instead of developing a general circuit model for analytical studies. In our manuscript, we preserve the closed-loop control dynamics in Thevenin equivalent circuit model for AC, DC, and DS subgrids. Then, we further elucidate how the secondary frequency/voltage control loops, the controllers of two ILCs, and the proposed novel full-timescale dynamic concatenator, can be uniformly incorporated into the GECM. It is worth noting that all the components in GECM are represented by transfer functions, which means the derived GECM can interpret system dynamics from low to high frequency range. For clarity, a comparison between this article and related researches is given in Table. \ref{comparison}.
	
	\textit{3) The system stability consideration}: For the hybrid AC/DC/DS MG configuration studied in this paper, system stability can be affected by the loading conditions and the cutoff frequency $\omega_0$ of the dynamic concatenator. For the loading conditions, the capacities of the individual subgrids determine whether the system can maintain stable operation. If the steady-state power allocated to a subgrid exceeds its capacity, the overall system may become unstable. For the selection of $\omega_0$, as shown in Fig. 7, given that the DSP sampling frequency requirement is satisfied, increasing $\omega_0$ will reduce the high-frequency bandwidth of the dynamic concatenator. This mainly affects the transient dynamic performance of the system and does not cause instability. The experimental results in Section V further demonstrate the stable operation of the hybrid AC/DC/DS MG under the selected parameters and operating conditions.

	\begin{table*}[t]
	\centering
	\caption{A Comparison Between the Proposed GECM and the State-of-the-Art Researches}
	\label{comparison}
	
	\footnotesize
	\setlength{\tabcolsep}{2pt}

	\begin{tabular}{
			>{\centering\arraybackslash}m{1.25cm}
			>{\centering\arraybackslash}m{2.5cm}
			>{\centering\arraybackslash}m{2cm}
			>{\centering\arraybackslash}m{2cm}
			>{\centering\arraybackslash}m{2.5cm}
			>{\centering\arraybackslash}m{1.6cm}
			>{\centering\arraybackslash}m{1.6cm}
			>{\centering\arraybackslash}m{1.8cm}
		}
		
		\toprule
		\textbf{Study} &
		\textbf{HESS With Complementary Dynamic Responses} &
		\textbf{Transient Inertia Sharing} &
		\textbf{Static global Power Sharing} &
		\textbf{Full-timescale dynamic concatenation of transient inertia sharing and GPS} &
		\textbf{Global equivalent circuit representation} &
		\textbf{Secondary frequency and voltage restoration} &
		\textbf{HESS and ILC dynamics included in GECM} \\
		\midrule
		
		\cite{ZhangZhe2021} &
		$\times$ & $\times$ & $\checkmark$ &
		$\times$ & $\times$ & $\times$ & $\times$ \\
		\midrule
		
		\cite{LinPengfeng2025TIE} &
		$\checkmark$ & $\checkmark$ & $\checkmark$ &
		$\times$ & $\checkmark$ & $\times$ & $\checkmark$ \\
		\midrule
		
		\cite{ZhangZhe2023} &
		$\times$ & $\checkmark$ & $\times$ &
		$\times$ & $\times$ & $\times$ & $\times$ \\
		\midrule
		
		\cite{WangJie} &
		$\times$ & $\checkmark$ & $\times$ &
		$\times$ & $\times$ & $\times$ & $\times$ \\
		\midrule
		
		\cite{WangPeng} &
		$\times$ & $\times$ & $\checkmark$ &
		$\times$ & $\times$ & $\times$ & $\times$ \\
		\midrule
		
		\cite{JinChi2018} &
		$\times$ & $\times$ & $\checkmark$ &
		$\times$ & $\times$ & $\checkmark$ & $\times$ \\
		\midrule
		
		This article &
		$\checkmark$ & $\checkmark$ & $\checkmark$ &
		$\checkmark$ & $\checkmark$ & $\checkmark$ & $\checkmark$ \\
		\bottomrule
		
	\end{tabular}
\end{table*}

	\begin{table}[t] 
		\centering
		\renewcommand{\arraystretch}{1}
		\setlength{\tabcolsep}{3pt}
		\caption{ Key System Parameter Settings}
		\label{system_parameters}
		\small
		\begin{tabular}{>{\centering\arraybackslash}m{2cm} 
				>{\centering\arraybackslash}m{4cm} 
				>{\centering\arraybackslash}m{2.5cm}}
			\toprule
			\textbf{System Components} & \textbf{Parameter} & \textbf{Setting} \\
			\midrule
			\multirow{4}{*}{\textbf{AC subgrid}} 
			& Input and output voltage & 380 V, 710 V \\
			& Maximum frequency  & 51 Hz \\
			& Minimum frequency  & 49 Hz \\
			& Nominal frequency  & 50 Hz \\
			& Maximum output power & 20 kW \\
			& Inertia and damping coefficient & $H_{ac}= 2$, $D_{ac}= 1$ \\
			\midrule
			\multirow{4}{*}{\textbf{DC subgrid}} 
			& Input voltage & 100 V \\
			& Maximum DC bus voltage  & 380 V \\
			& Minimum DC bus voltage  & 370 V \\
			& Nominal DC bus voltage  & 375 V \\
			& Maximum output power & 20 kW \\
			& Inertia and damping coefficient & $H_{dc}= 3$, $D_{dc}= 1$ \\
			\midrule
			\multirow{4}{*}{\textbf{DS subgrid}} 
			& Input voltage & 250 V \\
			& Maximum DS bus voltage  & 710 V \\
			& Minimum DS bus voltage  & 690 V \\
			& Nominal DS bus voltage  & 700 V \\
			& Maximum output power & 20 kW \\
			& Traditional and integral droop coefficients & $y_H= 7.5$, $y_L= 35.5$ \\
			\midrule
			\textbf{Other parameters of virtual inertia} 
			& $T_G, F_{HP}, T_{CH}, T_{RH}$ & 0.1, 0.3, 0.2, 7 \\
			\bottomrule
		\end{tabular}
		\vspace{-1.5em}
	\end{table}
	
	\section{ EXPERIMENTAL VERIFICATION} %Section IV
	
\begin{figure}[t]
		\setlength{\abovecaptionskip}{-3pt}
		\setlength{\belowcaptionskip}{-5pt}
		\centering
		\includegraphics[scale=1]{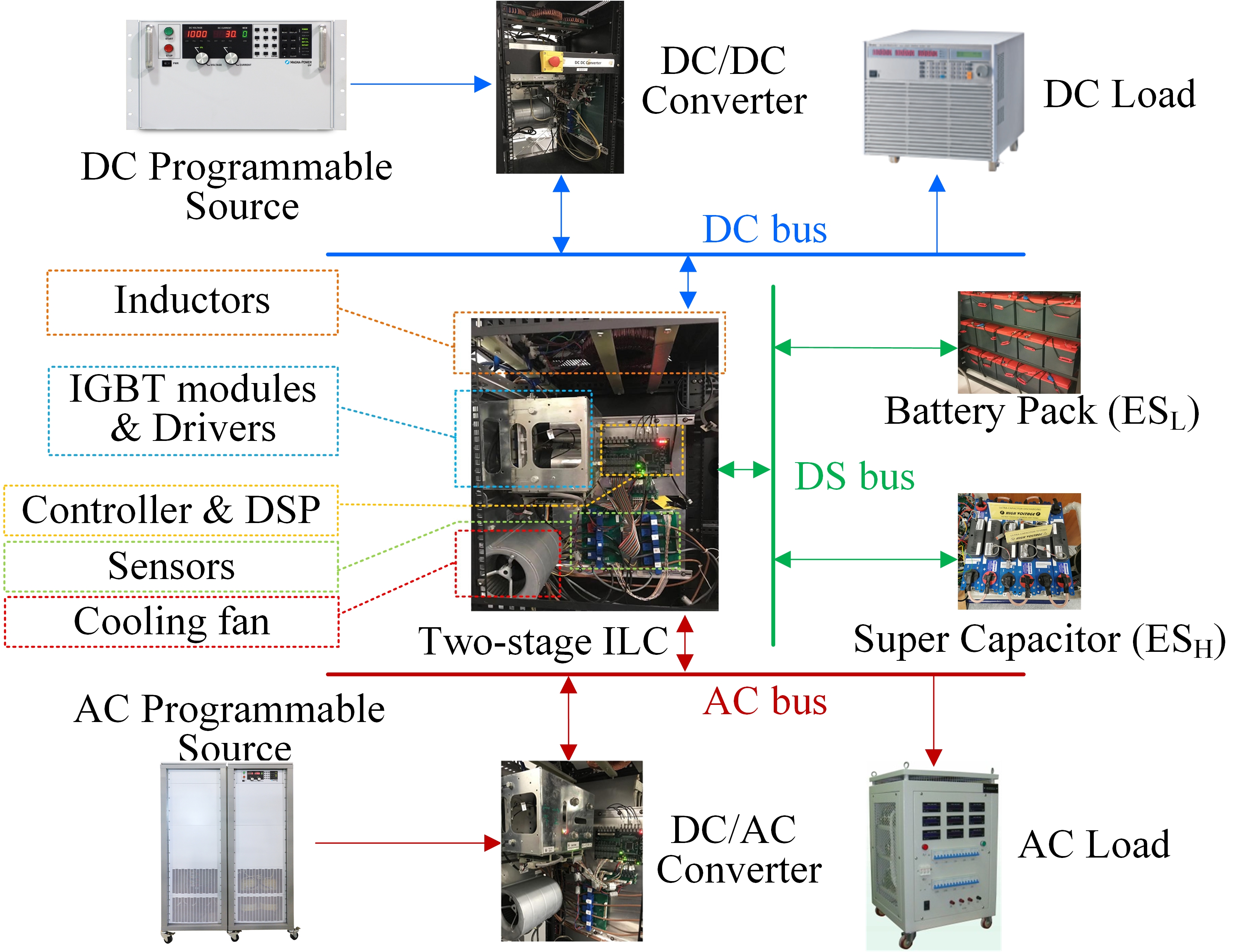}
		\caption{ Experimental platform.}
		\label{Fig. experiment}
		\vspace{-1.5em}
	\end{figure}

	According to the system configuration shown in Fig. \ref{Fig. AC/DC/DS MG}, experiments shown in Fig. \ref{Fig. experiment} were conducted in a hybrid MG to validate the effectiveness of  proposed full-timescale power management and steady-state frequency/voltage autonomous restoration control strategy. TABLE \ref{system_parameters} displays the system parameter configurations. The control algorithms were performed on a digital signal processor (DSP-TMS320F28335) together with a controller board. The DC bus was set up by a programmable DC power supply (Chroma 62100H-600S) via a droop-characterized DC/DC converter. The DC power supply (Magna-Power MTD1000-100/380) formed an AC subgrid through a DC/AC converter. $\rm{ES_L}$ and $\rm{ES_H}$ formed HESS and were connected to the DC capacitor of the two-stage ILC (a.k.a. DS bus). The power ratings of battery pack and super-capacitor pack are 14.4 kW and 5.6 kW. 
	
	Experimental waveforms were captured by a Tektronix MDO3000 Mixed Domain Oscilloscope and subsequently exported to MATLAB for further analysis of the system’s transient response metrics. The RoCoV and RoCoF at the instant of a load disturbance were approximated by differentiating the voltage or frequency measurements at the disturbance instant and the immediately succeeding sampling point. In essence, $\Delta x^*(t)/P_{ox}^*(t)|{_{t \to {0^+}}}$ represents the rates of change of frequency and voltage (RoCoF / RoCoV) at the instant of a load disturbance. Leveraging the proposed GECM, the rates of change of p.u. AC frequency, DC voltage, and DS voltage after inertia transfer can be quantitatively calculated by (\ref{eq. Initial value theorem with FTS}). Performing the load disturbance, AC subgrid RoCoF, DC subgrid RoCoV, and DS subgrid RoCoV are computed as in Table \ref{Corresponding global inertia coefficient and subgrid’s RoCoF/RoCoV for various ID coefficients}.
	
\begin{table}[t]
	\centering
	\renewcommand{\arraystretch}{1.1}
	\setlength{\tabcolsep}{3pt}
	\caption{global inertia coefficient $H_G$ and subgrid’s RoCoF/RoCoV for various ID coefficients $y_H$ by GECM of (\ref{eq. Initial value theorem with FTS}). (Note: DC, AC, and DS subgrids undergo load variations of 14 kW, 12 kW, and 10 kW.) }
	\label{Corresponding global inertia coefficient and subgrid’s RoCoF/RoCoV for various ID coefficients}
	\small
	\begin{tabular}{c c c c c}
		\toprule
		\makecell{\textbf{$y_H$}} &
		\makecell{\textbf{ $H_G$}} &
		\makecell{\textbf{RoCoF}\\ \textbf{of AC subgrid}} &
		\makecell{\textbf{RoCoV}\\ \textbf{of DC subgrid}} &
		\makecell{\textbf{RoCoV}\\ \textbf{of DS subgrid}} \\
		\midrule
		15 & 12.5/3 & 3.67 Hz/s & 27.36 V/s & 51.12 V/s \\
		30 & 20/3   & 2.29 Hz/s & 17.10 V/s & 31.95 V/s \\
		\bottomrule
	\end{tabular}
	\vspace{-1.0em}
\end{table}

    \vspace{-1.0em}
	\subsection{Hybrid Microgrid with Inertia Transfer but Without Frequency/Voltage Restoration Control}
	To assess the influence of the FTS dynamic concatenator and the frequency/voltage restoration control on the hybrid AC/DC/DS MG,  Fig. \ref{Fig. response isolated} (a) depicts the dynamic responses of the hybrid AC/DC/DS MG with only transient inertia transfer are reported in \cite{LinPengfeng2025TIE, ZhangZhe2023}, while Fig. \ref{Fig. response isolated} (b) illustrates that DC, AC, and DS subgrids undergo load variations of 14 kW, 12 kW, and 10 kW at first. Upon the settlement of the initial  perturbation, the load power of AC subgrid is subsequently raised by an additional 6 kW. As illustrated in  Fig. \ref{Fig. response isolated} (a), the RoCoF of the AC subgrid and the RoCoV of DC and DS subgrids at the moment of the load disturbance are 27.32 V/s, 3.66 Hz/s, and 51.08 V/s, respectively, which aligns closely with the values reported in Table \ref{Corresponding global inertia coefficient and subgrid’s RoCoF/RoCoV for various ID coefficients}. This concordance substantiates the capability of the global ECM to accurately quantify the system’s effective inertia.
	
	Owing to the absence of frequency and voltage restoration mechanisms, both quantities drift away from their nominal values following load disturbances. Under successive disturbances, the resulting frequency and voltage excursions may exceed permissible thresholds, thereby jeopardizing the secure and reliable operation of the hybrid MG. In addition, since the transient inertia transfer mechanism is fundamentally oriented toward enhancing the system’s dynamic responsiveness while neglecting its steady-state power allocation, subgrids fail to attain a proportionate distribution of global system load in steady state, as illustrated in Fig. \ref{Fig. response isolated}(b). This situation may drive certain subgrids beyond their permissible operating limits, whereas the available capacity margins of others remain substantially underutilized.
	
	\vspace{-1em}
	\subsection{Hybrid Microgrid with FTS Dynamic Concatenator and Frequency/Voltage Restoration Control}
\begin{figure}[t]
		\setlength{\abovecaptionskip}{-3pt}
		\setlength{\belowcaptionskip}{-5pt}
		\centering
		\includegraphics[scale=1]{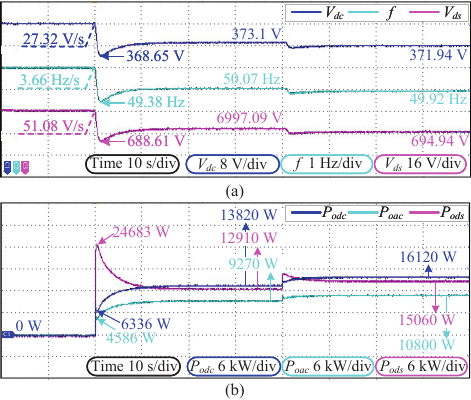}
		\caption{Response of hybrid MG with inertia transfer but Without frequency/voltage restorations. (a) DC bus voltage and AC frequency response. (b) Output power of DC and AC subgrid.}
		\label{Fig. response isolated}
		\vspace{-1.0em}
	\end{figure}

\begin{figure}[t]
		\setlength{\abovecaptionskip}{-3pt}
		\setlength{\belowcaptionskip}{-5pt}
		\centering
		\includegraphics[scale=1]{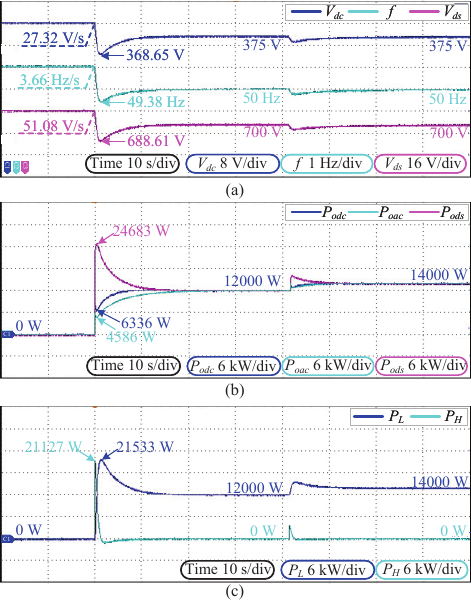}
		\caption{Response process of hybrid MG with FTS dynamic concatenator and frequency/voltage restoration control ($y_H$=15). (a) DC bus voltage and AC frequency response. (b) The output power of DC, AC and DS subgrid. (c) Power allocation of $\rm{ES_L}$ and $\rm{ES_H}$ in DS subgrid.}
		\label{Fig. response with ID=15}
		\vspace{-1.5em}
	\end{figure}

	Fig. \ref{Fig. response with ID=15} illustrates the dynamic responses of hybrid MG with the FTS dynamic concatenator and the frequency/voltage restoration control. A comparison between  Fig. \ref{Fig. response isolated}(a) and Fig.  \ref{Fig. response with ID=15}(a) reveals that the steady-state autonomous restoration control effectively maintains the DC bus voltage and AC frequency at their nominal values despite repeated load disturbances. 
	
	Fig. \ref{Fig. response with ID=15}(b) illustrates that the DC, AC, and DS subgrids undergo load perturbations of 14 kW, 12 kW, and 10 kW during the initial disturbance, and subsequently stabilize at an identical steady-state output of 16 kW. When an additional 6 kW load is subsequently applied to the AC subgrid, the system maintains GPS. This result demonstrates that ILC equipped with the FTS dynamic concatenator can perform inertia sharing control during the transient phase and progressively transition to steady-state GPS  control, thereby simultaneously ensuring favorable transient response performance and optimized steady-state power allocation.
	
	The FTS dynamic concatenator is designed to coordinate transient and steady-state power regulation across different subgrids. Its primary role is to ensure a smooth transition between transient inertia transfer and steady-state GPS, rather than introducing additional dynamic modes into the system. From the results presented in Fig. \ref{Fig. response with ID=15}, it can be observed that the proposed control framework maintains stable operation under loading condition. The dynamic responses of AC frequency, DC and DS bus voltages, remain well regulated without exhibiting oscillatory or divergent behavior. This indicates that the incorporation of the FTS dynamic concatenator does not compromise the overall stability of the system.

	Fig. \ref{Fig. response with ID=15}(c) illustrates the output power of the $\rm{E{S_L}}$ and $\rm{E{S_H}}$ in HESS, where $\rm{E{S_L}}$ and $\rm{E{S_H}}$ actively handle low frequency and high frequency power,	respectively. When the load change occurs, $\rm{E{S_H}}$ immediately responds to the rapid change in load power and then gradually decreases to zero, while the output power of $\rm{E{S_L}}$ slowly rises until it takes over all the load of DS subgrid. This result indicates that the coordinated control under conventional P–V and ID droop control strategies can effectively leverage the respective advantages of high-power-density and high-energy-density energy storage systems.

	\vspace{-1em}
	\subsection{Increase the Integral Droop Coefficient $y_H$ in DS Subgrid to Further Drive Transient Inertia Sharing}
	
	\begin{figure}[t]
		\setlength{\abovecaptionskip}{-3pt}
		\setlength{\belowcaptionskip}{-5pt}
		\centering
		\includegraphics[scale=1]{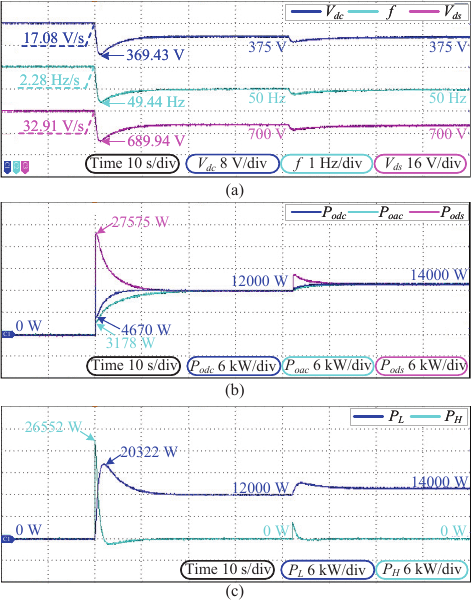}
		\caption{Response process of hybrid MG with FTS dynamic concatenator and frequency/voltage restoration control. ($y_H$=30) (a) DC bus voltage and AC frequency response. (b) The output power of DC, AC and DS subgrid. (c) Power allocation of $\rm{ES_L}$ and $\rm{ES_H}$ in DS subgrid.}
		\label{Fig. response with ID=30}
		\vspace{-1.5em}
	\end{figure}
	
	Fig. \ref{Fig. response with ID=30} shows the experimental results of the hybrid AC/DC/DS MG where $y_H$ is increased to 30. The transient response of DC bus voltage and AC frequency are further optimized. As shown in this figure, the RoCoV of the DC subgrid and the RoCoF of the AC subgrid are reduced to 17.08 V/s and 2.28 Hz/s, respectively, while the nadir of the DC bus voltage and the minimum AC frequency are improved to 369.43 V and 49.44 Hz, respectively. This is because the additional inertia of the DS subgrid is transferred to other low-inertia subgrids through inertia-sharing control, thereby enhancing their transient response characteristics.
	
	In Fig. \ref{Fig. response with ID=30} (b), it can be observed that the increase of $y_H$ does not affect the static global power sharing. Regarding the transient process, the transient power peak of ds subgrid rises to 27.575 kW, while the peak output power of dc and ac subgrids decreases relatively. Owing to the elevated transient power of the DS subgrid and the increase of $y_H$, the $\rm{ES_H}$ compensates for a greater proportion of the high-frequency components, whereas the $\rm{ES_L}$ exhibits a more gradual and stable output as shown in Fig. \ref{Fig. response with ID=30}(c). The results demonstrate that increasing ID coefficient of $\rm{ES_H}$ enables DS subgrid to contribute more effectively to system inertia sharing, thereby alleviating transient power surges in other subgrids. Moreover, a higher $y_H$ improves the DS subgrid’s high/low frequency power decomposition. As a result, the transient stress of $\rm{ES_L}$ is reduced, which in turn prolongs the operational lifespan of the hybrid energy storage system in DS subgrid.

	\section{Conclusion}
	This paper addresses the challenge of FTS coordinated regulation in hybrid AC/DC/DS MGs. To this end, an FTS power scheduling strategy for hybrid AC/DC/DS MGs with dynamic concatenation and autonomous frequency/voltage restoration is presented, together with an analytical model that characterizes the system’s full-timescale response dynamics. Based on the developed analytical model, simulation, and hardware-in-the-loop experimental results, the proposed strategy is demonstrated to be effective. The FTS dynamic concatenator enables transient inertia transfer among subgrids, thereby enhancing the overall dynamic performance and facilitating a smooth transition toward steady-state global power sharing, ensuring efficient utilization of subgrid capacity. Meanwhile, the autonomous frequency/voltage restoration control successfully eliminates the steady-state deviations induced by conventional virtual inertia, allowing voltages/frequency to autonomously converge to their nominal values, which prevents from severe steady-state offsets that may threaten critical loads or trigger cascading faults. Furthermore, the proposed GECM clearly captures the full-timescale dynamic behavior and inter-subgrid inertia propagation mechanism, providing a novel analytical perspective  for coordinated global inertia transfer control.

	\vspace{-0.5em}
	\bibliographystyle{IEEEtran}
	\bibliography{reference}

@ARTICLE{YangFang,
  author={Yang, Fang and Feng, Xianyong and Li, Zhao},
  journal={IEEE Transactions on Industry Applications}, 
  title={Advanced Microgrid Energy Management System for Future Sustainable and Resilient Power Grid}, 
  year={2019},
  volume={55},
  number={6},
  pages={7251-7260},
  doi={10.1109/TIA.2019.2912133}
}

@ARTICLE{Zia,
  author={Zia, Muhammad Fahad and Elbouchikhi, Elhoussin and Benbouzid, Mohamed and Guerrero, Josep M.},
  journal={IEEE Transactions on Industry Applications}, 
  title={Energy Management System for an Islanded Microgrid With Convex Relaxation}, 
  year={2019},
  volume={55},
  number={6},
  pages={7175-7185},
  doi={10.1109/TIA.2019.2917357}}

@article{Blaabjerg,
   author={Blaabjerg, F. and Teodorescu, R. and Liserre, M. and Timbus, A.V.},
  journal={IEEE Transactions on Industrial Electronics}, 
  title={Overview of Control and Grid Synchronization for Distributed Power Generation Systems}, 
  year={2006},
  volume={53},
  number={5},
  pages={1398-1409},
  doi={10.1109/TIE.2006.881997}
}

@article{JinChi2014,
 author={Jin, Chi and Wang, Peng and Xiao, Jianfang and Tang, Yi and Choo, Fook Hoong},
  journal={IEEE Transactions on Industrial Electronics}, 
  title={Implementation of Hierarchical Control in DC Microgrids}, 
  year={2014},
  volume={61},
  number={8},
  pages={4032-4042},
  doi={10.1109/TIE.2013.2286563}
}

@article{ZhangZhe2021,
  author={Zhang, Zhe and Jin, Chi and Tang, Yi and Dong, Chaoyu and Lin, Pengfeng and Mi, Yang and Wang, Peng},
  journal={IEEE Transactions on Industrial Electronics}, 
  title={A Modulized Three-Port Interlinking Converter for Hybrid AC/DC/DS Microgrids Featured With a Decentralized Power Management Strategy}, 
  year={2021},
  volume={68},
  number={12},
  pages={12430-12440},
  doi={10.1109/TIE.2020.3040660}
}

@article{LiFeng,
  author={Li, Feng and Zhu, Jiebei and Yu, Lujie and Bu, Siqi and Zhao, Haoran and Zhao, Junbo and Xu, Yan and Guerrero, Josep M. and Wang, Chengshan},
  journal={IEEE Transactions on Smart Grid}, 
  title={An Imbalance-Status-Enabled Autonomous Global Power-Sharing Scheme for Solid-State Transformer Interconnected Hybrid AC/DC Microgrids}, 
  year={2023},
  volume={14},
  number={3},
  pages={1750-1762},
  doi={10.1109/TSG.2022.3216853}
}

@article{Loh,
  author={Loh, Poh Chiang and Li, Ding and Chai, Yi Kang and Blaabjerg, Frede},
  journal={IEEE Transactions on Power Electronics}, 
  title={Autonomous Operation of Hybrid Microgrid With AC and DC Subgrids}, 
  year={2013},
  volume={28},
  number={5},
  pages={2214-2223},
  doi={10.1109/TPEL.2012.2214792}}

@ARTICLE{Lee,
  author={Lee, Gyu-Sub and Kwon, Do-Hoon and Moon, Seung-Il and Hwang, Pyeong-Ik},
  journal={IEEE Transactions on Power Systems}, 
  title={A Coordinated Control Strategy for LCC HVDC Systems for Frequency Support with Suppression of AC Voltage Fluctuations}, 
  year={2020},
  volume={35},
  number={4},
  pages={2804-2815},
  doi={10.1109/TPWRS.2020.2964336}}

@ARTICLE{ChenJianbo,
  author={Chen, Jianbo and Yue, Dong and Dou, Chunxia and Chen, Lei and Weng, Shengxuan and Li, Yanman},
  journal={IEEE Transactions on Industrial Informatics}, 
  title={A Virtual Complex Impedance Based $P-\dot{V}$ Droop Method for Parallel-Connected Inverters in Low-Voltage AC Microgrids}, 
  year={2021},
  volume={17},
  number={3},
  pages={1763-1773},
  doi={10.1109/TII.2020.2997054}}

@article{LohPoh,
  author={Loh, Poh Chiang and Li, Ding and Chai, Yi Kang and F.{Blaabjerg}},
  journal={IEEE Transactions on Industry Applications}, 
  title={Autonomous Control of Interlinking Converter With Energy Storage in Hybrid AC–DC Microgrid}, 
  year={2013},
  volume={49},
  number={3},
  pages={1374-1382},
}

@ARTICLE{Jayan,
  author={Jayan, Vijesh and Ghias, Amer Mohammad Yusuf Mohammad},
  journal={IEEE Transactions on Power Electronics}, 
  title={Computationally-Efficient Model Predictive Control of Dual-Output Multilevel Converter in Hybrid Microgrid}, 
  year={2023},
  volume={38},
  number={5},
  pages={5898-5910},
  doi={10.1109/TPEL.2023.3239437}}

@article{LinPengfeng2019TSG,
  author={Lin, Pengfeng and Wang, Peng and Jin, Chi and Xiao, Jianfang and Li, Xiaoqiang and Guo, Fanghong and Zhang, Chuanlin},
  journal={IEEE Transactions on Smart Grid}, 
  title={A Distributed Power Management Strategy for Multi-Paralleled Bidirectional Interlinking Converters in Hybrid AC/DC Microgrids}, 
  year={2019},
  volume={10},
  number={5},
  pages={5696-5711},
  doi={10.1109/TSG.2018.2890420}
}

@ARTICLE{LinPengfeng2019TSG_2,
  author={Lin, Pengfeng and Jin, Chi and Xiao, Jianfang and Li, Xiaoqiang and Shi, Donghan and Tang, Yi and Wang, Peng},
  journal={IEEE Transactions on Smart Grid}, 
  title={A Distributed Control Architecture for Global System Economic Operation in Autonomous Hybrid AC/DC Microgrids}, 
  year={2019},
  volume={10},
  number={3},
  pages={2603-2617},
  doi={10.1109/TSG.2018.2805839}}

@article{C.L,
	author={C.L. {Sulzberger}},
  journal={IEEE Power and Energy Magazine}, 
  title={Triumph of AC. 2. The battle of the currents}, 
  year={2003},
  volume={1},
  number={4},
  pages={70-73},
  doi={10.1109/MPAE.2003.1213534}}

@article{HongQiteng,
  title={Addressing frequency control challenges in future low-inertia power systems: A Great Britain perspective},
  author={Hong, Qiteng and Khan, Md Asif Uddin and Henderson, Callum and Egea-{\`A}lvarez, Agust{\'\i} and Tzelepis, Dimitrios and Booth, Campbell},
  journal={Engineering},
  volume={7},
  number={8},
  pages={1057--1063},
  year={2021},
  publisher={Elsevier}
}

@article{LiChang,
 author={Li, Chang and Li, Yong and Cao, Yijia and Zhu, Hongqi and Rehtanz, Christian and Häger, Ulf},
  journal={IEEE Journal of Emerging and Selected Topics in Power Electronics}, 
  title={Virtual Synchronous Generator Control for Damping DC-Side Resonance of VSC-MTDC System}, 
  year={2018},
  volume={6},
  number={3},
  pages={1054-1064},
  doi={10.1109/JESTPE.2018.2827361}}

@TECHREPORT{IEC61727,
  institution = {International Electrotechnical Commission},
  title       = {Photovoltaic (PV) systems -- Characteristics of the utility interface},
  number      = {IEC 61727},
  year        = {2004},
  address     = {Geneva, Switzerland},
}

@article{LinPengfeng2025TIE,
  author  = {Pengfeng Lin and Qingzuo Meng and Miao Zhu and Frede Blaabjerg},
  title   = {Dynamic Circuit-Based Unified Power Regulation for Hybrid AC/DC/DS Microgrids: A Comprehensive Approach to Static and Transient Control},
  journal = {IEEE Transactions on Industrial Electronics},
  year    = {2025},
  note = { doi: 10.1109/TIE.2025.3585029}
}

@article{ZhangZhe2023,
  author={Zhang, Zhe and Fang, Jingyang and Dong, Chaoyu and Jin, Chi and Tang, Yi},
  journal={IEEE Transactions on Industrial Electronics}, 
  title={Enhanced Grid Frequency and DC-Link Voltage Regulation in Hybrid AC/DC Microgrids Through Bidirectional Virtual Inertia Support}, 
  year={2023},
  volume={70},
  number={7},
  pages={6931-6940},
  doi={10.1109/TIE.2022.3203757}
}

@article{WangJie,
  author={Wang, Jie and Huang, Wentao and Tai, Nengling and Yu, Moduo and Li, Ran and Zhang, Yong},
  journal={IEEE Transactions on Power Systems}, 
  title={A Bidirectional Virtual Inertia Control Strategy for the Interconnected Converter of Standalone AC/DC Hybrid Microgrids}, 
  year={2024},
  volume={39},
  number={1},
  pages={745-754},
  doi={10.1109/TPWRS.2023.3246522}
}

@article{WangPeng,
  author={Wang, Peng and Jin, Chi and Zhu, Dexuan and Tang, Yi and Loh, Poh Chiang and Choo, Fook Hoong},
  journal={IEEE Transactions on Industrial Electronics}, 
  title={Distributed Control for Autonomous Operation of a Three-Port AC/DC/DS Hybrid Microgrid}, 
  year={2015},
  volume={62},
  number={2},
  pages={1279-1290},
  doi={10.1109/TIE.2014.2347913}
}

@article{JinChi2018,
  author={Jin, Chi and Wang, Junjun and Wang, Peng},
  journal={CSEE Journal of Power and Energy Systems}, 
  title={Coordinated secondary control for autonomous hybrid three-port AC/DC/DS microgrid}, 
  year={2018},
  volume={4},
  number={1},
  pages={1-10},
  doi={10.17775/CSEEJPES.2016.01400}
}

@article{Kundur,
	title={Power system stability},
	author={P.{Kundur}},
	journal={Power system stability and control},
	volume={10},
	pages={7--1},
	year={2007},
	publisher={CRC Press New York}
}

@article{LinPengfeng2018TPE,
 author={Lin, Pengfeng and Wang, Peng and Xiao, Jianfang and Wang, Junjun and Jin, Chi and Tang, Yi},
  journal={IEEE Transactions on Power Electronics}, 
  title={An Integral Droop for Transient Power Allocation and Output Impedance Shaping of Hybrid Energy Storage System in DC Microgrid}, 
  year={2018},
  volume={33},
  number={7},
  pages={6262-6277},
  doi={10.1109/TPEL.2017.2741262}
}

@manual{TMS32_28335,
  title        = {TMS320F28335 Digital Signal Controller (DSC) Datasheet},
  author       = {{Texas Instruments}},
  organization = {Texas Instruments},
  address      = {Dallas, TX, USA},
  year         = {2022},
  number       = {SPRS439Q},
  note         = {[Online]. Available: \url{https://www.ti.com/lit/ds/symlink/tms320f28335.pdf}}
}

@ARTICLE{Luxiaonan,
  author={Dragičević, Tomislav and Lu, Xiaonan and Vasquez, Juan C. and Guerrero, Josep M.},
  journal={IEEE Transactions on Power Electronics}, 
  title={DC Microgrids—Part I: A Review of Control Strategies and Stabilization Techniques}, 
  year={2016},
  volume={31},
  number={7},
  pages={4876-4891},
  doi={10.1109/TPEL.2015.2478859}}

@ARTICLE{YangPengcheng,
  author={Yang, Pengcheng and Yu, Miao and Wu, Qiuwei and Hatziargyriou, Nikos and Xia, Yanghong and Wei, Wei},
  journal={IEEE Transactions on Smart Grid}, 
  title={Decentralized Bidirectional Voltage Supporting Control for Multi-Mode Hybrid AC/DC Microgrid}, 
  year={2020},
  volume={11},
  number={3},
  pages={2615-2626},
  }

	\begin{IEEEbiography}[{\includegraphics	[width=0.95in,height=1.25in,clip]{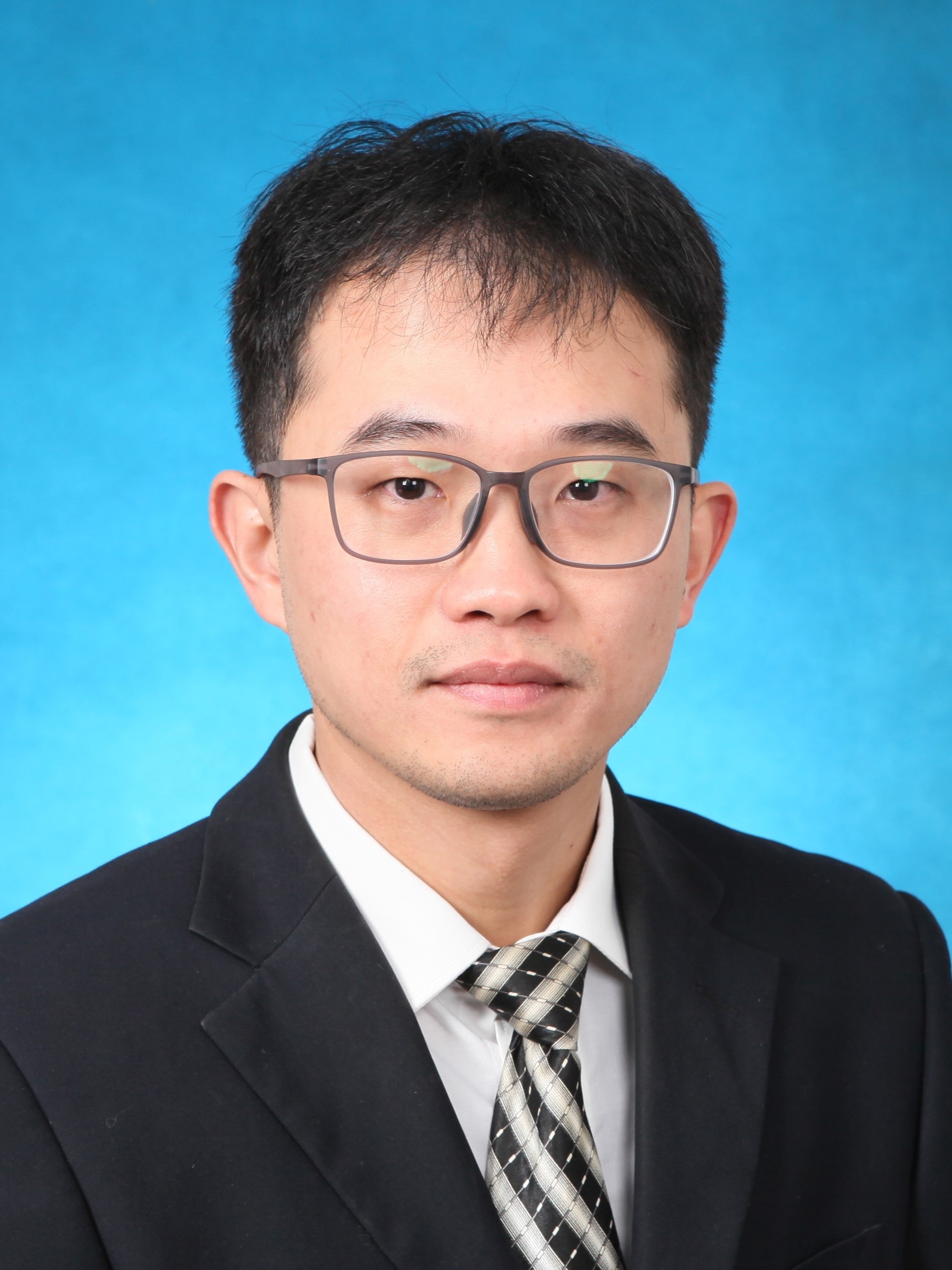}}]
	{Pengfeng Lin} (Senior Member, IEEE) received the B.S. and M.S. degree in electrical engineering from Southwest Jiaotong University, China, in 2013 and 2015 respectively, and the Ph.D. degree in Interdisciplinary Graduate Programme from Nanyang Technological Universty, Singapore, in 2020. He was with Energy Research Institute @ NTU as a Research Fellow and with Électricité de France (EDF) as a Smart Grid Energy Expert / Research Scientist. He is in the global top 2\% of Scientists released by Standard University. He is currently a Marie Skłodowska-Curie Actions (MSCA) Fellow at the University of Cambridge and an Associate Professor at Shanghai Jiao Tong University. His research interests include power system stability / reliability analyses, smart grid cyber-physical security, and artificial intelligence analytic.
	\end{IEEEbiography}
	
	%\vspace{-0.5mm}
	\begin{IEEEbiography}[{\includegraphics	[width=0.95in,height=1.25in,clip]{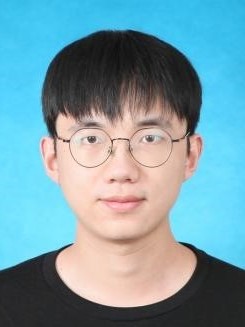}}]
	{Qingzuo MENG} (Student Member, IEEE)  received the B.E. and M.S. degree in electrical engineering from  Shanghai Jiao Tong University (SJTU), Shanghai, China, in 2023 and 2026 respectively. He is currently working at the Electric Power Dispatching and Control Center of Guizhou Power Grid.
	\end{IEEEbiography}
	
	\begin{IEEEbiography}[{\includegraphics	[width=0.95in,height=1.25in,clip]{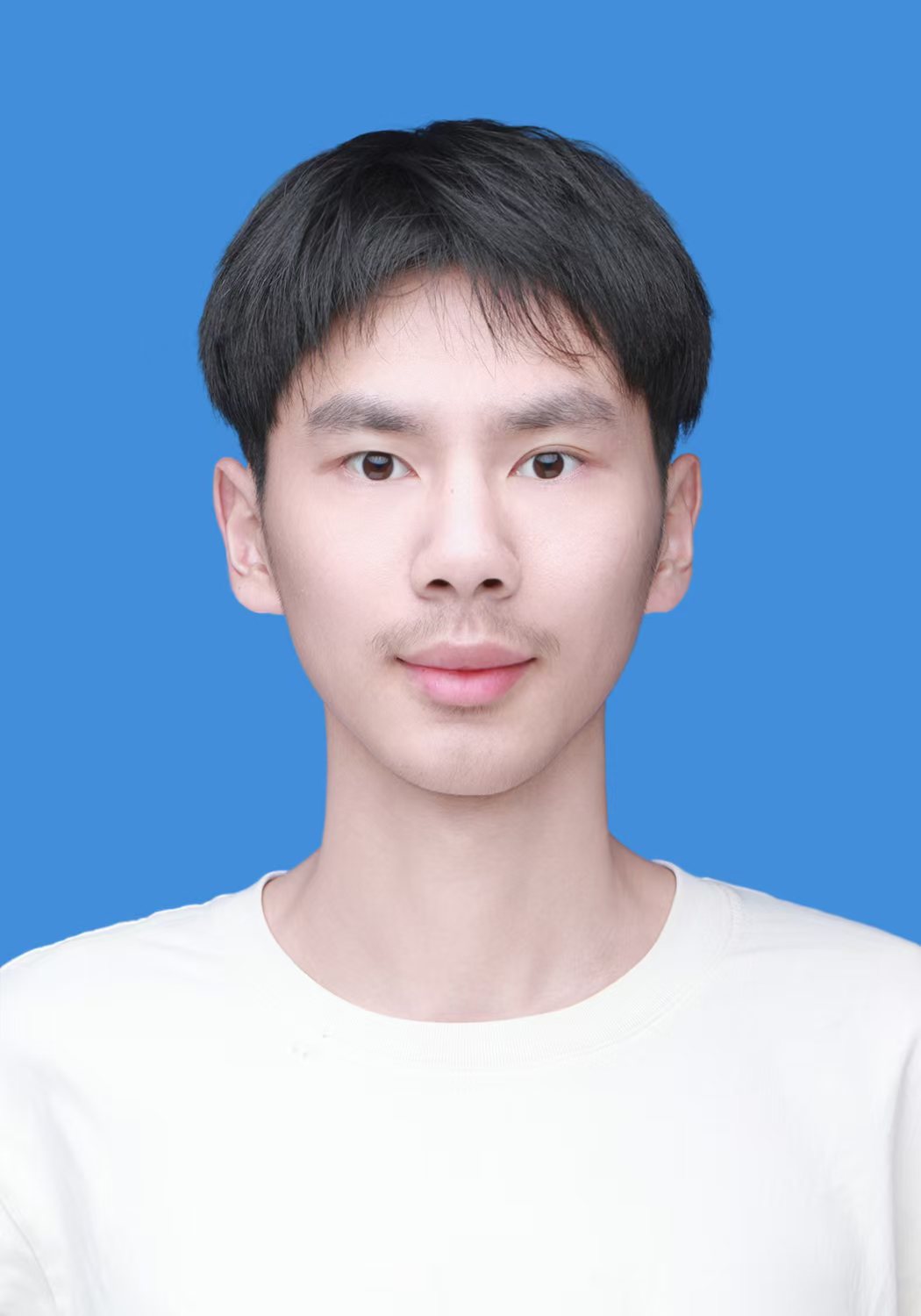}}]
	{Yujie Wang} (Student Member, IEEE)  received the B.E. degree in electrical engineering   from  Shanghai Jiao Tong University (SJTU), Shanghai, China, in 2025.  He is currently working toward the M.S. degree in electrical engineering of SJTU. 
	\end{IEEEbiography}

	\begin{IEEEbiography}[{\includegraphics	[width=0.95in,height=1.2in,clip]{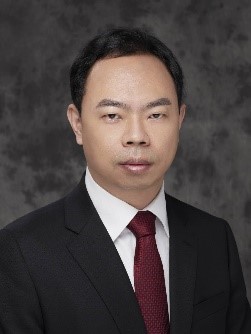}}]
	{Miao Zhu} (Senior member, IEEE)  received the B.Sc. degree from Southeast University, China, in 2001, and the Ph.D. degree from Nanyang Technological University, Singapore, in 2009, both in electrical engineering. From 2008 to 2009, he was with Meiden Asia Pte Ltd, Singapore, as an R\&D Engineer. After that, he had been a Scientist in the Experimental Power Grid Centre (EPGC), A*STAR, Singapore. In Aug. 2012, he joined Shanghai Jiao Tong University (SJTU), China, as Research Professor with the honor title of Distinguished Researcher. Currently, he is the Distinguished Professor of Electrical Engineering Department, SJTU. 
	\end{IEEEbiography}

	\begin{IEEEbiography}[{\includegraphics	[width=0.95in,height=1.25in,clip]{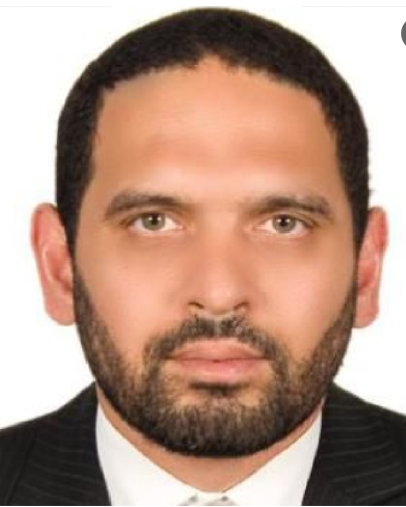}}]
	{Amer M. Ghias} (Senior member, IEEE) received the B.Sc. degree in Electrical Engineering from Saint Cloud State University, USA, in 2001, the M.Eng. degree in Telecommunications from the University of Limerick, Ireland, in 2006, and the Ph.D. degree in Electrical Engineering from the University of New South Wales (UNSW), Australia, in 2014. He has over seven years of industry experience in Kuwait, where he worked as an Electrical Engineer, Project Engineer, and Project Manager before transitioning to academia. He held academic positions at UNSW Australia and the University of Sharjah, UAE, and later served as an Assistant Professor at Nanyang Technological University (NTU), Singapore, where he was also the Cluster Director of Power Electronics at the Energy Research Institute @ NTU (ERI@N). He is currently a Senior Lecturer (Associate Professor) at The University of Glasgow, UK. He serves as an Associate Editor for IET Power Electronics and is a Senior Member of IEEE.
	\end{IEEEbiography}
	
	\begin{IEEEbiography}[{\includegraphics	[width=0.95in,height=1.25in,clip]{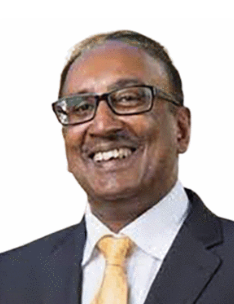}}]
	{Syed M. Islam} (Fellow, IEEE ) received his B.Sc., M.Sc., and Ph.D. degrees in Electrical Power Engineering in 1979, 1983, and 1988, respectively. He is the Executive Dean and a Professor at the School of Science, Engineering, and Information Technology, Federation University Australia, Ballarat, Melbourne. He serves as the Chair of the Australasian Committee for Power Engineering (ACPE), is a member of the Steering Committee of the Australian Power Institute, and holds Fellowships with Engineers Australia, IEEE IAS, PES, DEIS, and IET. He is an Editor for IEEE Transactions on Sustainable Energy and a Distinguished Lecturer for IEEE PES, regularly reviewing submissions for various IEEE Transactions.
	\end{IEEEbiography}

	\begin{IEEEbiography}[{\includegraphics	[width=0.95in,height=1.3in,clip]{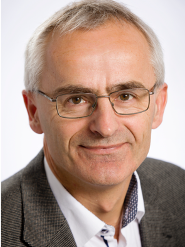}}]
	{Frede Blaabjerg} (Fellow, IEEE) was with ABB-Scandia, Randers, Denmark, from 1987 to 1988. From 1988 to 1992, he got the PhD degree in Electrical Engineering at Aalborg University in 1995. He became an Assistant Professor in 1992, an Associate Professor in 1996, and a Full Professor of power electronics and drives in 1998. From 2017 he became a Villum Investigator. He is Honoris Causa at University Politehnica Timisoara (UPT), Romania and Tallinn Technical University (TTU) in Estonia.
	\end{IEEEbiography}

	\newpage

	\begin{comment}
		\section{Biography Section}
		If you have an EPS/PDF photo (graphicx package needed), extra braces are
		needed around the contents of the optional argument to biography to prevent
		the LaTeX parser from getting confused when it sees the complicated
		$\backslash${\tt{includegraphics}} command within an optional argument. (You can create
		your own custom macro containing the $\backslash${\tt{includegraphics}} command to make things
		simpler here.)
		
		\vspace{11pt}
		
		\bf{If you include a photo:}\vspace{-33pt}
		\begin{IEEEbiography}[{\includegraphics[width=1in,height=1.25in,clip,keepaspectratio]{fig1}}]{Michael Shell}
			Use $\backslash${\tt{begin\{IEEEbiography\}}} and then for the 1st argument use $\backslash${\tt{includegraphics}} to declare and link the author photo.
			Use the author name as the 3rd argument followed by the biography text.
		\end{IEEEbiography}
		
		\vspace{11pt}
		
		\bf{If you will not include a photo:}\vspace{-33pt}
		\begin{IEEEbiographynophoto}{John Doe}
			Use $\backslash${\tt{begin\{IEEEbiographynophoto\}}} and the author name as the argument followed by the biography text.
		\end{IEEEbiographynophoto}
	\end{comment}

	\vfill
	
\end{document}